\documentclass{aa}  

\usepackage{graphicx}
\usepackage{txfonts}
\usepackage{subcaption}         % necessary for continued figures, example in section 3
\usepackage{lscape}             % to rotate a single page table, example in appendix.
\usepackage{placeins}           % useful with \FloatBarrier, to keep 
\usepackage{hyperref}
\usepackage[normalem]{ulem}
\usepackage{color}

\newcommand{\fiddy}{SGAS-J1050}
\newcommand{\elf}{SGAS-J1110}
\newcommand{\batleth}{SGAS-J1429}
\newcommand{\geese}{SGAS-J1527}
\newcommand{\eye}{Cosmic Eye}

\newcommand{\oiii}{[\textrm{O}\textsc{iii}]}

\newcommand{\oiiiauroral}{[\textrm{O}\textsc{iii}]\ensuremath{\lambda4364}}
\newcommand{\oiiiuvdbl}{\textrm{O}\textsc{iii}]\ensuremath{\lambda\lambda1660,1666}}

\newcommand{\oii}{[\textrm{O}\textsc{ii}]}
\newcommand{\oiibrightdbl}{[\textrm{O}\textsc{ii}]\ensuremath{\lambda\lambda3727,3730}}
\newcommand{\oiiauroraldbl}{[\textrm{O}\textsc{ii}]\ensuremath{\lambda\lambda7322,7332}}

\newcommand{\oi}{[\textrm{O}\textsc{i}]}

\newcommand{\sii}{[\textrm{S}\textsc{ii}]}
\newcommand{\siidbl}{[\textrm{S}\textsc{ii}]\ensuremath{\lambda\lambda6718,6733}}
\newcommand{\siiiauroral}{[\textrm{S}\textsc{iii}]\ensuremath{\lambda6314}}

\newcommand{\siii}{[\textrm{S}\textsc{iii}]}

\newcommand{\siiib}{[\textrm{S}\textsc{iii}]\ensuremath{\lambda9532}}

\newcommand{\nii}{[\textrm{N}\textsc{ii}]}
\newcommand{\niiauroral}{[\textrm{N}\textsc{ii}]\ensuremath{\lambda5756}}

\newcommand{\niiiuv}{\textrm{N}\textsc{iii}]\ensuremath{\lambda1750}}

\newcommand{\arivdbl}{[\textrm{Ar}\textsc{iv}]\ensuremath{\lambda\lambda4713,4741}}
\newcommand{\ariva}{[\textrm{Ar}\textsc{iv}]\ensuremath{\lambda4713}}

\newcommand{\neiii}{[\textrm{Ne}\textsc{iii}]}

\newcommand{\neiiia}{[\textrm{Ne}\textsc{iii}]\ensuremath{\lambda3870}}
\newcommand{\neiiib}{[\textrm{Ne}\textsc{iii}]\ensuremath{\lambda3969}}

\newcommand{\hi}{\textrm{H}\textsc{i}}
\newcommand{\hii}{\textrm{H}\textsc{ii}}
\newcommand{\halpha}{\textrm{H}\ensuremath{\alpha}}
\newcommand{\hbeta}{\textrm{H}\ensuremath{\beta}}

\newcommand{\hepsilon}{\textrm{H}\ensuremath{\epsilon}}

\newcommand{\hei}{\textrm{He}\textsc{i}}
\newcommand{\heii}{\textrm{He}\textsc{ii}}
\newcommand{\heiiopt}{\textrm{He}\textsc{ii}\ensuremath{\lambda4686}}
\newcommand{\heiiuv}{\textrm{He}\textsc{ii}\ensuremath{\lambda1640}}
\newcommand{\heia}{\textrm{He}\textsc{i}\ensuremath{\lambda4471}}
\newcommand{\heibeiargon}{\textrm{He}\textsc{i}\ensuremath{\lambda4714}}

\newcommand{\siliiidbl}{\textrm{Si}\textsc{iii}]\ensuremath{\lambda\lambda1882,1892}}

\newcommand{\ciiidbl}{\textrm{C}\textsc{iii}]\ensuremath{\lambda\lambda1907,1909}}

\begin{document}

   \title{LEGGOS: Direct abundances of N, O, Ne, S, and Ar in five lensed galaxies at Cosmic Noon}

   \author{Brian Welch\inst{1}\corrauth{bwelch.astro@gmail.com}        % use \corrauth for the corresponding author
        \and Gourav Khullar\inst{2,3,4} %ਗੌਰਵ ਖੁੱਲਰ
        \and Taylor A. Hutchison\inst{5,6,7}
        \and Keren Sharon\inst{8}
        \and Pedram Abedi\inst{8}
        \and Matthew B. Bayliss\inst{9}
        \and Michael Florian\inst{10,11}
        \and Dylan Berry\inst{2}
        \and Jacqueline Antwi-Danso\inst{12}
        \and Nikko J. Cleri\inst{13,14,15}
        \and H{\aa}kon Dahle\inst{16}
        \and Aleena Ebey Panzer\inst{9}
        \and Michael D. Gladders\inst{17,18}
        \and Rion Oh\inst{2,19} %오리온
        \and Cole Panzer\inst{9}
        \and Jane R. Rigby\inst{5}
        \and T.\ Emil Rivera-Thorsen\inst{20}
        }
        % ADD OTHERS TOO

   \institute{International Space Science Institute, Hallerstrasse 6, 3012 Bern, Switzerland
   \and Department of Astronomy \& the DiRAC Institute, University of Washington, Physics-Astronomy Building, Box 351580, Seattle, WA 98195-1700, USA
   \and eScience Institute, University of Washington, Physics-Astronomy Building, Box 351580, Seattle, WA 98195-1700, USA
   \and Department of Physics and Astronomy and PITT PACC, University of Pittsburgh, Pittsburgh, PA 15260, USA
   \and Astrophysics Science Division, Code 660, NASA Goddard Space Flight Center, 8800 Greenbelt Rd., Greenbelt, MD 20771, USA
   \and Department of Astronomy, University of Maryland, Baltimore County, MD 21250, USA
   \and Center for Research and Exploration in Space Science and Technology, NASA/GSFC, Greenbelt, MD 20771 USA
   \and Department of Astronomy, University of Michigan, 1085 S. University Ave, Ann Arbor, MI 48109, USA
   \and Department of Physics, University of Cincinnati, Cincinnati, OH 45221, USA
   \and Steward Observatory, University of Arizona, 933 North Cherry Avenue, Tucson, AZ 85721, USA
   \and Eureka Scientific, 2452 Delmer Street Suite 100 Oakland, CA 94602-3017
   \and David A. Dunlap Dept. of Astronomy and Astrophysics, University of Toronto, 50 St. George Street, Toronto, M5S 3H4, Canada
   \and Department of Astronomy and Astrophysics, The Pennsylvania State University, University Park, PA 16802, USA
   \and Institute for Computational \& Data Sciences, The Pennsylvania State University, University Park, PA 16802, USA
   \and Institute for Gravitation and the Cosmos, The Pennsylvania State University, University Park, PA 16802, USA
   \and Institute of Theoretical Astrophysics, University of Oslo, P.O. Box 1029, Blindern, NO-0315 Oslo, Norway
   \and Department of Astronomy and Astrophysics, University of Chicago, 5640 South Ellis Avenue, Chicago, IL 60637, USA
   \and Kavli Institute for Cosmological Physics, University of Chicago, 5640 South Ellis Avenue, Chicago, IL 60637, USA
   \and Department of Physics, KAIST, Daejeon 34141, Republic of Korea
   \and The Oskar Klein Centre, Department of Astronomy, Stockholm University, AlbaNova 10691, Stockholm, Sweden
   }

   \date{Received ---}

  \abstract {
   We present direct $T_e$ chemical abundances in five strongly lensed galaxies at Cosmic Noon ($2.481 \leq z \leq 3.625$) using JWST/NIRSpec data from the LEGGOS survey. We measure gas-phase abundances of N, O, S, and Ar in all five galaxies, and Ne in two galaxies. Three galaxies have electron temperature constraints from multiple different ionization zones, and we find that these are broadly consistent with temperature scaling relations observed in both local and high-$z$ galaxies. We find a range of oxygen abundances $8.04 \leq 12+\log(\text{O/H}) \leq 8.79$ (22 -- 126\% $Z_{\odot}$). The ratios of N/O and Ne/O are consistent with trends observed in local galaxies. We do not observe any evidence for significant N enhancement in our sample, though the youngest galaxy in our sample (SGAS-J1050) has a mildly elevated $\log(\text{N/O}) = 1.20 \pm 0.08$, which we suggest may be driven by a population of young massive stars. The ratios of S/O and Ar/O are generally sub-solar, similar to trends observed in other high-$z$ galaxies. We find that the S/O and Ar/O abundances are sub-solar, consistent with enrichment from core-collapse supernovae (CCSNe). Modeling of the star formation histories of the LEGGOS galaxies supports CCSNe enrichment, as each galaxy shows a recent period of star formation lasting $\lesssim 100$ Myr, indicating that type Ia supernovae would not yet have had enough time to contribute significantly to the gas-phase abundances of these galaxies. 
   }


   \maketitle
\nolinenumbers

%%%%%%%%%%%%%%%%%%%%%%%%%%%%%%%%%%%%%%%%%%%%%%%%%%%%%%%%%%%%%%
\section{Introduction}

The chemical abundance patterns of galaxies provide a powerful tool to study their formation histories. Elements are synthesized in stars and distributed throughout the interstellar medium (ISM) via stellar winds and supernova explosions, and successive generations of stars progressively build up the chemical abundances of galaxies \citep[e.g.,][]{burbidge_synthesis_1957,kobayashi_nucleosynthesis_2025}. 

It is possible to measure the abundances of elements in gaseous nebulae within galaxies using the relative strengths of nebular emission lines \citep[e.g.,][]{osterbrock_astrophysics_2006}. One of the more precise methods for making such measurements in nebulae outside the Milky Way is the ``direct $T_e$" method, which utilizes a combination of emission line ratios to calculate the electron temperature $T_e$ and electron density $n_e$. The abundances of elements can then be inferred using the temperature, density, and strength of bright collisionally-excited lines \citep[CELs; e.g.][]{dinerstein_abundances_1990,peimbert_nebular_2017,kewley_understanding_2019,curti_chemical_2025}. The $T_e$ method is often limited by the faintness of the key temperature-sensitive auroral lines, which are often just a few percent the flux of \hbeta\ and therefore require high signal-to-noise ratio (SNR) spectra to detect.

The direct $T_e$ method has been employed in many studies of the low-redshift ($z$) universe, where the faint auroral lines are more accessible. Large samples of multi-element $T_e$ abundances have been assembled using large spectroscopic surveys \citep[e.g.,][]{izotov_chemical_2006,scholte_electron_2026}. Additionally, individual \hii\ regions in nearby galaxies have been studied using both compilations of slit spectra \citep[e.g.,][]{berg_chaos_2015,berg_chaos_2020,rogers_chaos_2021} and by defining \hii\ regions from integral field unit (IFU) spectra \citep[e.g.,][]{kreckel_mapping_2019,kreckel_temperature_2025}. Meanwhile, other studies have used IFU spectra to create spatially-resolved abundance maps of galaxies, finding significant variations in gas-phase abundances within galaxies \citep[e.g.,][]{menacho_ionized_2021,rickards_vaught_interstellar_2025}.

Use of the direct $T_e$ method in distant galaxies has increased significantly in recent years thanks to the sensitivity and wavelength coverage of JWST \citep{gardner_james_2023,rigby_science_2023}. JWST observations have enabled direct $T_e$ abundances to be measured in larger samples of galaxies at cosmic noon than was possible with previous instruments \citep{stanton_jwst_2025,bhattacharya_unveiling_2025,rogers_cecilia_2026}. These data also enable $T_e$ abundance measurements out to higher redshifts than was possible with previous observatories \citep[e.g.,][]{arellano-cordova_first_2022,schaerer_first_2022,taylor_metallicities_2022,brinchmann_high-z_2023,cameron_nitrogen_2023,curti_chemical_2023,katz_first_2023,rhoads_finding_2023,trump_physical_2023,senchyna_gn-z11_2024,marques-chaves_extreme_2024,castellano_jwst_2024,hsiao_jwst_2024,sanders_direct_2024,sanders_aurora_2025,arellano-cordova_jwst_2025,marques-chaves_signatures_2026,arellano-cordova_self-consistent_2026,berg_fleeting_2026}.

One interesting finding from these high-$z$ studies is the existence of objects with elevated nitrogen abundance relative to their low oxygen abundance \citep[e.g.,][]{cameron_nitrogen_2023,senchyna_gn-z11_2024,marques-chaves_extreme_2024,castellano_jwst_2024,schaerer_discovery_2024,welch_sunburst_2025,curti_marta_2025,berg_fleeting_2026,ji_connecting_2026,naidu_cosmic_2026,morel_discovery_2025}. The mechanisms driving these elevated N/O abundances remain debated, with suggested sources including AGB stars in bursty star forming environments \citep[e.g.][]{dantona_gn-z11_2023,mcclymont_thesan-zoom_2025}, short-term enrichment from Wolf-Rayet (WR) stars \citep[e.g.][]{kobayashi_rapid_2024,rivera-thorsen_sunburst_2024,welch_sunburst_2025,curti_marta_2025,berg_fleeting_2026} or very massive stars \citep[VMS, e.g.][]{vink_very_2023}, and more exotic objects such as supermassive stars \citep[e.g.][]{charbonnel_n-enhancement_2023}.

Recent work has begun investigating whether excess N is a common feature at high-$z$, finding mixed results. \cite{cataldi_tracing_2025} find a significant excess of N/O on average in galaxies at $z>1$, meanwhile \cite{schaerer_nitrogen_2026} find no evidence of N enhancement on average in Lyman-continuum emitting galaxies at $z\sim 3$, and \cite{cameron_jades_2026} find a modest increase in the average N/O at low O/H across $1.5 < z < 7$, though a handful of galaxies have larger N excesses. Using strong-line diagnostics, \cite{morel_discovery_2025} find that the fraction of galaxies with high-ionization N emission and related N/O excesses increases towards higher redshift. With these varied results, the sources of potential N enhancement across high-$z$ galaxy populations remain unclear. 

Another unexpected trend is that some studies have found evidence for systematic underabundances of the $\alpha$-elements argon and sulfur relative to oxygen \citep{rogers_cecilia_2024,bhattacharya_unveiling_2025,stanton_jwst_2025,rogers_cecilia_2026}. However other observations have found S/O and Ar/O abundances in line with expectations from local galaxies \citep{welch_templates_2024,bhattacharya_unveiling_2025,welch_sunburst_2025}. The enrichment mechanism often proposed to explain these relative abundance trends is enrichment from core-collapse supernovae (CCSNe) without additional contribution from type Ia supernovae (SNIa), since SNIa contribute significant amounts of S and Ar without corresponding O enrichment \citep{kobayashi_nucleosynthesis_2025}. The S and Ar abundance ratio is therefore sensitive to the star formation history of the galaxy. 

In this paper, we explore the multi-element abundance patterns of five strongly lensed galaxies from the LEnsing and Galaxy Growth: Observing Substructures (LEGGOS) survey \citep{khullar_leggos_2026}. Here we present the spatially integrated abundances, while follow-up work (Welch et al., in prep.) will explore variations between \hii\ regions within these lensed galaxies. 

This paper is structured as follows. We discuss the data used in this analysis in Section \ref{sec:data}, and we explain our emission line fitting procedure in Section \ref{sec:emissionlines}. We describe the spectral energy distribution (SED) fitting process in Section \ref{sec:sedfits}. We present our temperature, density, and chemical abundance measurements in Section \ref{sec:physconditions}, and discuss the results of this analysis in Section \ref{sec:discussion}. Finally, we present our conclusions in Section \ref{sec:conclusions}.

%%%%%%%%%%%%%%%%%%%%%%%%%%%%%%%%%%%%%%%%%%%%%%%%%%%%%%%%%%%%%%
\section{Sample Description \& Data Reduction} \label{sec:data}

\subsection{Sample}
The galaxies analyzed in this work come from the LEGGOS program (PID 0425, PIs Florian, Khullar; PID 03843, PI Bayliss), which observed six strongly lensed galaxies with both NIRCam imaging and NIRSpec/IFS spectroscopy. LEGGOS also included two archival targets, SGAS-J1226 from the TEMPLATES Early Release Science (ERS) program \citep[PID 01355, PIs Rigby, Viera;][]{rigby_jwst_2025} and the Sunburst Arc \citep[PID 02555, PI Rivera-Thorsen;][]{rivera-thorsen_sunburst_2025}. 

We select five galaxies from the 8 LEGGOS targets for this work. We do not analyze SGAS-J2111 and SGAS-J1226 due to a lack of detectable $T_e$-sensitive auroral emission lines \citep{welch_templates_2024}. Though the Sunburst Arc has detected auroral lines \citep{welch_sunburst_2025}, several of these lines are partially cut off by the NIRSpec detector gap. We therefore do not analyze the full-arc spectrum to avoid systematic biases driven by the incomplete coverage of the key auroral lines. 

\subsection{JWST Data Reduction}

The data collection and data reduction are described in detail in \cite{khullar_leggos_2026}. We briefly summarize the key points here. 

\subsubsection{JWST/NIRSpec}

Each target was observed with the JWST/NIRSpec Integral Field Spectrograph (IFS) \citep{boker_near-infrared_2022,boker_-orbit_2023}. All targets have data in the G235M medium-resolution ($R\sim 1000$) grating, and the low-resolution ($R\sim 100$) prism as part of GO-0425. \elf\ was additionally observed with the G140H and G235H high-resolution ($R\sim 3000$) gratings as part of GO-03843. For this analysis, we include the G235H data from both the initial observation of \elf, which was affected by a mirror tilt event \citep[see][for details]{khullar_leggos_2026}, and the subsequent re-observations. While the PSF of the initial tilt-affected data is likely skewed relative to the other observations, this change has minimal impact on the spatially-integrated spectrum. 

NIRSpec data are reduced using the JWST data reduction pipeline version 1.20.2 \citep{bushouse_jwst_2025}, with calibration reference data set (CRDS) \texttt{jwst1466.pmap}. We utilize the NSClean algorithm \citep{rauscher_nsclean_2024} via the \texttt{clean\_flicker\_noise} step of the JWST pipeline to remove 1/f noise, and we employ the \texttt{baryon-sweep} post-processing tool to remove any remaining outliers and artifacts in the final data cubes \citep{hutchison_templates_2024}. 

For medium and high resolution spectra, we subtract the predicted background model generated by the JWST Background Tool (JBT\footnote{\href{https://jwst-docs.stsci.edu/jwst-other-tools/jwst-backgrounds-tool}{https://jwst-docs.stsci.edu/jwst-other-tools/jwst-backgrounds-tool}}). The JBT has a known issue in which the backgrounds are overestimated below $\lambda < 1.2 \mu\text{m}$ \citep{rigby_how_2023}. We therefore measure the background using off-target spaxels within the IFS field of view. As described in \cite{khullar_leggos_2026}, the background measured using this process is consistent with the backgrounds measured in the empty fixed slits, indicating that the background is not contaminated by residual light from the source. 

For this work, we re-drizzle the final NIRSpec data cubes to a common pixel grid for each target using the \texttt{nspax\_x} and \texttt{nspax\_y} keywords of the \texttt{spec3} pipeline's cube-building step. The number of spaxels in the x and y directions are selected to ensure that the full IFS FoV is contained in the final cube for each target. 

We also manually align the output data cubes to the NIRCam WCS solution using an offset file passed to the cube-building step of the \texttt{spec3} pipeline. The NIRSpec IFS WCS solutions are imprecise at up to a few tenths of an arcsecond; the FoV is too small to contain additional stars to be used as reference objects, so they rely exclusively on the telescope's guiding system for the final WCS solution. We measure the WCS offsets between NIRSpec and NIRCam using bright continuum features present in each IFS cube. For all targets, this includes several bright clumps identified visually in each source. For the Cosmic Eye, we additionally use the centroid of the central lens galaxy as a reference point when defining the WCS shift. 

Shifting the WCS to be aligned to NIRCam and drizzling the output data cubes to an identical coordinate grid ensures that our spectral extraction regions (defined below in Section \ref{subsec:ext1d}) are identical between different disperser settings, which may include offsets in the position angle based on the observation timing. 

Finally, we correct the uncertainty array for each NIRSpec cube based on the observed fluctuations in blank areas of the IFS FoV. We use a method similar to that described in \cite{rivera-thorsen_sunburst_2025}, in which we take the standard deviation of pixel values in blank regions at each wavelength slice and use the greater of this standard deviation and the pipeline-calculated uncertainty as our final uncertainty spectrum. 

\begin{figure*}
    \centering
    \includegraphics[width=\textwidth]{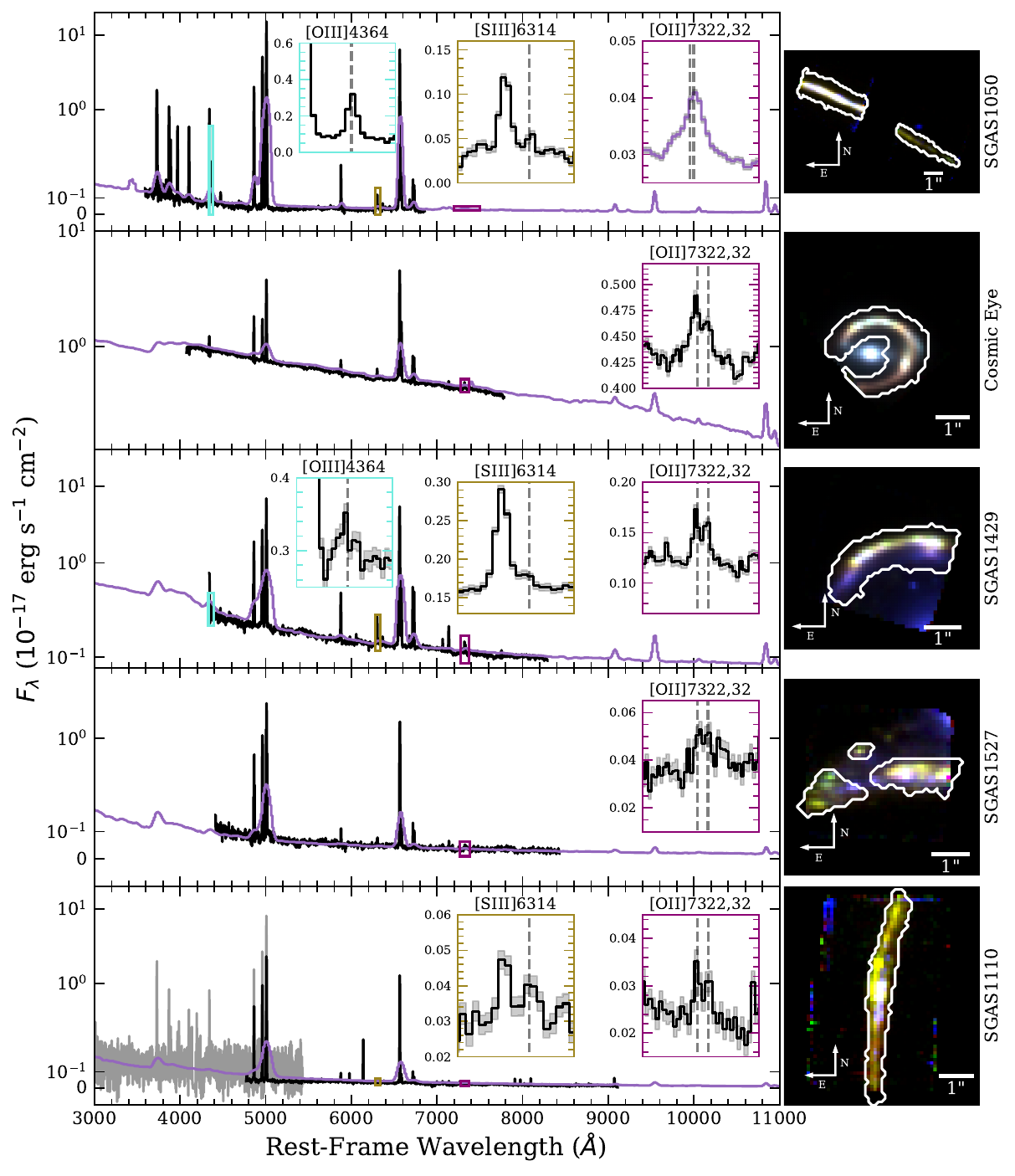}
    \caption{Extracted spatially integrated 1D spectra are shown for the galaxies analyzed in this work. G235M spectra are shown in black and prism spectra are shown in purple in each panel. The G140H spectrum for \elf\ is shown in grey. Detected temperature-sensitive auroral lines are highlighted in inset panels for each spectrum. All targets have \oiiauroraldbl\ detections, three have \siiiauroral\ detections, and two have \oiiiauroral\ detections at SNR $>3$. The color images in the right column are composites of the \halpha\ (red), \oiii\ (green), and continuum (blue) emission from the NIRSpec IFU data cubes. The white contour indicates the aperture used to extract the 1D spectra. }
    \label{fig:spectra}
\end{figure*}

\subsubsection{JWST/NIRCam}

Each target was observed with JWST/NIRCam \citep{rieke_performance_2023}. Observation descriptions and data reduction methods are reported in \cite{khullar_leggos_2026}. Briefly, the NIRCam data were WCS-aligned to archival HST imaging and drizzled to a 0\farcs03 pixel grid. Residual 1/f noise was removed between the Level 2 and Level 3 pipelines using a custom method described in \cite{khullar_leggos_2026}. 

\subsection{Archival ground-based spectroscopy} \label{subsec:grounddata}

We include archival rest-UV spectroscopy from several sources to complement our JWST rest-optical spectra. For three of our targets (\eye, \geese, and \batleth), rest-UV spectra were obtained as part of the MEGaSaURA program \citep{rigby_magellan_2018}. These data were obtained with the Magellan/MagE spectrograph \citep{marshall_mage_2008}, and data reduction is detailed in \cite{rigby_magellan_2018}. \batleth\ and the Cosmic Eye both have only a single spatially coadded spectrum from the MEGaSaURA data; however, \geese\ has two spectra of two distinct regions in the galaxy. Because the NIRSpec IFU data covers the full source image, we choose to stack these two spectra. 

\eye\ has Keck/ESI spectroscopy published in \cite{quider_study_2010}, which we analyze in addition to the MagE spectrum for this target.

\fiddy\ was observed with the MagE \citep{marshall_mage_2008}, IMACS, and FIRE \citep{simcoe_fire_2013} spectrographs on Magellan, as well as with the GMOS spectrograph on Gemini \citep{hook_gemini-north_2004}, as detailed in \cite{bayliss_physical_2014}. For the present analysis, we use the published emission line fluxes reported in \cite{bayliss_physical_2014}. 

\elf\ was observed with the GMOS spectrograph on Gemini \citep{hook_gemini-north_2004}, and the Blue Channel spectrograph on the MMT, as detailed in \cite{johnson_star_2017-1}. We use the same data reduction described in \cite{johnson_star_2017-1} for the present analysis.

\subsection{Lens models}

We utilize updated gravitational lens models of each system in our sample based on JWST imaging and spectroscopy. The lens modeling methodology is based on that in \cite{sharon_strong_2020} and \cite{abedi_leggos_2026}, and uses the \texttt{lenstool} software \citep{jullo_bayesian_2007,jullo_multiscale_2009}. Additional details of these JWST-based lens models will be presented in a future publication. Here we briefly summarize the key points for each strong lens system. 

The lens model for \fiddy\ uses spectroscopically confirmed multiple image constraints from \cite{bayliss_physical_2014}, with the positions of multiply imaged clumps within these spectroscopically confirmed arcs updated based on the higher spatial resolution imaging from NIRCam. This yields a total of 48 multiple image constraints from three spectroscopically confirmed arcs. 

The \eye\ is primarily a galaxy-galaxy lens system, which has previously been modeled by \cite{dye_separation_2007}, though it is in the vicinity of a larger cluster-scale lens \citep{zitrin_strong-lensing_2016}. We model this system as a galaxy-galaxy lens with an external shear component to account for the effect of the nearby cluster lens. A total of 20 multiple image constraints are identified based on the NIRCam and NIRSpec IFU observations. 

\batleth\ was previously modeled by \cite{catan_molecular_2024}, but for this work we use a new model based on NIRCam-selected multiple images. We use a total of 31 constraints from two sources, one of which (the primary arc analyzed here) has a spectroscopic redshift. The redshift of the second source is left as a free parameter when optimizing the lens model. 

\geese\ is a complex lens system, as the primary arc analyzed here is primarily affected by the lensing potential of an elliptical galaxy 14\farcs5 from the core of the lensing cluster. This system has previously been modeled in \cite{sharon_strong_2020} and \cite{bordoloi_resolving_2022}. For this analysis, we use a total of 19 secure multiple image constraints from the primary arc defined based on NIRCam and NIRSpec data to update the galaxy-scale lensing potential, while the cluster-scale lens model is held fixed from previously published models. 

The JWST-based lens model for \elf\ is based on 88 constraints from 4 sources, one of which (the primary arc) has a spectroscopic redshift. The model is presented in detail in \cite{abedi_leggos_2026}. 

Table \ref{tab:sample} reports magnifications for each lensed galaxy. These values are the total magnification of the region used to extract spectra and photometry, not the magnification of the full arc. The magnifications reported here may differ from the total arc magnifications reported in previous studies \citep[e.g.,][]{abedi_leggos_2026}.

%%%%%%%%%%%%%%%%%%%%%%%%%%%%%%%%%%%%%%%%%%%%%%%%%%%%%%%%%%%%%%
\section{Emission Line Measurements} \label{sec:emissionlines}

\subsection{Extraction of 1D Spectra} \label{subsec:ext1d}

We extract spatially-integrated 1D spectra from an aperture including all spaxels with \halpha\ emission in the G235M grating detected at SNR $>5$.
The apertures and extracted spectra are shown in Figure \ref{fig:spectra}. 

For this step, we define SNR as follows.
In each spaxel, we first sum the flux density in a wavelength range around the emission line. 
We define the width of this region through visual inspection, yielding a width of  6 resolution elements for medium-resolution spectra.
We then sum the continuum in an equally-sized region adjacent to the emission line, selected visually to avoid any other nearby emission features. 
We subtract this continuum from the line flux to remove contamination from nearby foreground objects, for example, the lens galaxy present in the \eye\ IFS FoV. 
Next we sum the uncertainties in both the emission line region and the continuum region in quadrature, and use this combined uncertainty to calculate the total line SNR in each spaxel. 

For some targets (most prominently the \eye), the small PA and position offsets between G235M and prism observations mean that some regions of the lensed arc are only covered in a single pointing. We therefore remove any spaxels where one or more observations contain no data to mitigate any flux offsets or biases between the different dispersers.

We sum the flux density in each spaxel contained within our final mask, and sum the uncertainties from each spaxel in quadrature to create the spatially-integrated 1D spectrum. Figure \ref{fig:spectra} shows the resultant 1D spectra.

We correct for continuum emission from foreground lens galaxies in the vicinity of our targets after the initial spectrum has been extracted. To do so, we first measure the foreground galaxy spectrum by visually defining an aperture in the IFU cube. For the \eye, this aperture is at the center of the foreground galaxy. For \batleth, the foreground galaxy is slightly outside the IFU FoV, so we select continuum-bright spaxels near the edge of the FoV that are free from detector artifacts, which tend to be more common along the edges of the IFU. We then create a continuum image in the prism IFU cubes by summing along the wavelength axis. We fit the foreground galaxy in this continuum image using a 2D Sersic profile. We use this model to calculate the ratio of foreground galaxy flux in the target arc aperture to the foreground aperture, and we scale the foreground galaxy spectrum by this ratio. We smooth the scaled foreground spectrum using a smoothing spline, and subtract the smoothed foreground from the target arc spectrum. This procedure assumes that the foreground galaxy spectrum does not vary spatially. 

%For our analysis, we stack the G235M data with the G235H data for SGAS1110, downsampling onto the medium-resolution wavelength grid using the \texttt{spectres} package \citep{carnall_spectres_2017}. This process maximizes the SNR of the faint \oiiauroraldbl\ and \siiiauroral\ lines that fall within the G235 gratings. 
% IDK if this stays or not - I ended up using the G235M fluxes anyway, so maybe just stick with that? 

%We perform two checks to ensure consistency of the source flux between disperser settings. First, we find that the continuum level is consistent between the medium-resolution gratings and the prism, as shown in Figure \ref{fig:spectra}. The high-resolution G140H grating observed for SGAS1110 is an exception, since the high-resolution grating does not have significant continuum detections per wavelength slice. We also check for consistency of the measured flux for the \oiiia+\oiiib+\hbeta\ emission line group. We treat this group of lines collectively because they are blended in the prism data. We find that the summed line fluxes are consistent within $\sim 10\%$ between the different disperser settings.
% could add this back in, but it feels superflouous - the plot already shows that things look consistent. I could always add the O3+Hb flux to a table somewhere if I really need to make the point later, but I think its fine

%\red{ADD HERE - magnification corrections??}

\subsection{Emission Line Fitting} \label{subsec:linefits}
We fit emission lines following the methodology described in \cite{welch_templates_2024}. 
We first subtract the continuum by masking emission lines and smoothing the remaining spectrum using a boxcar convolution, with a boxcar size of 100\AA. 
The smoothed continuum spectrum is interpolated across masked emission line regions and then subtracted from the original spectrum. 

We use Gaussian profiles to fit emission lines in the continuum-subtracted spectra.
Line widths are fixed based on the NIRSpec dispersion files provided on JDox\footnote{\href{https://jwst-docs.stsci.edu/jwst-near-infrared-spectrograph/nirspec-instrumentation/nirspec-dispersers-and-filters}{https://jwst-docs.stsci.edu/jwst-near-infrared-spectrograph/nirspec-instrumentation/nirspec-dispersers-and-filters}}. 
%\red{Maybe update to the Shajib+ dispersions at some point? Or not because that paper doesn't include the prism?}
Line centers are allowed to vary within 2 resolution elements of the redshifted vacuum wavelength. 

\begin{figure}
    \centering
    \includegraphics[width=0.99\linewidth]{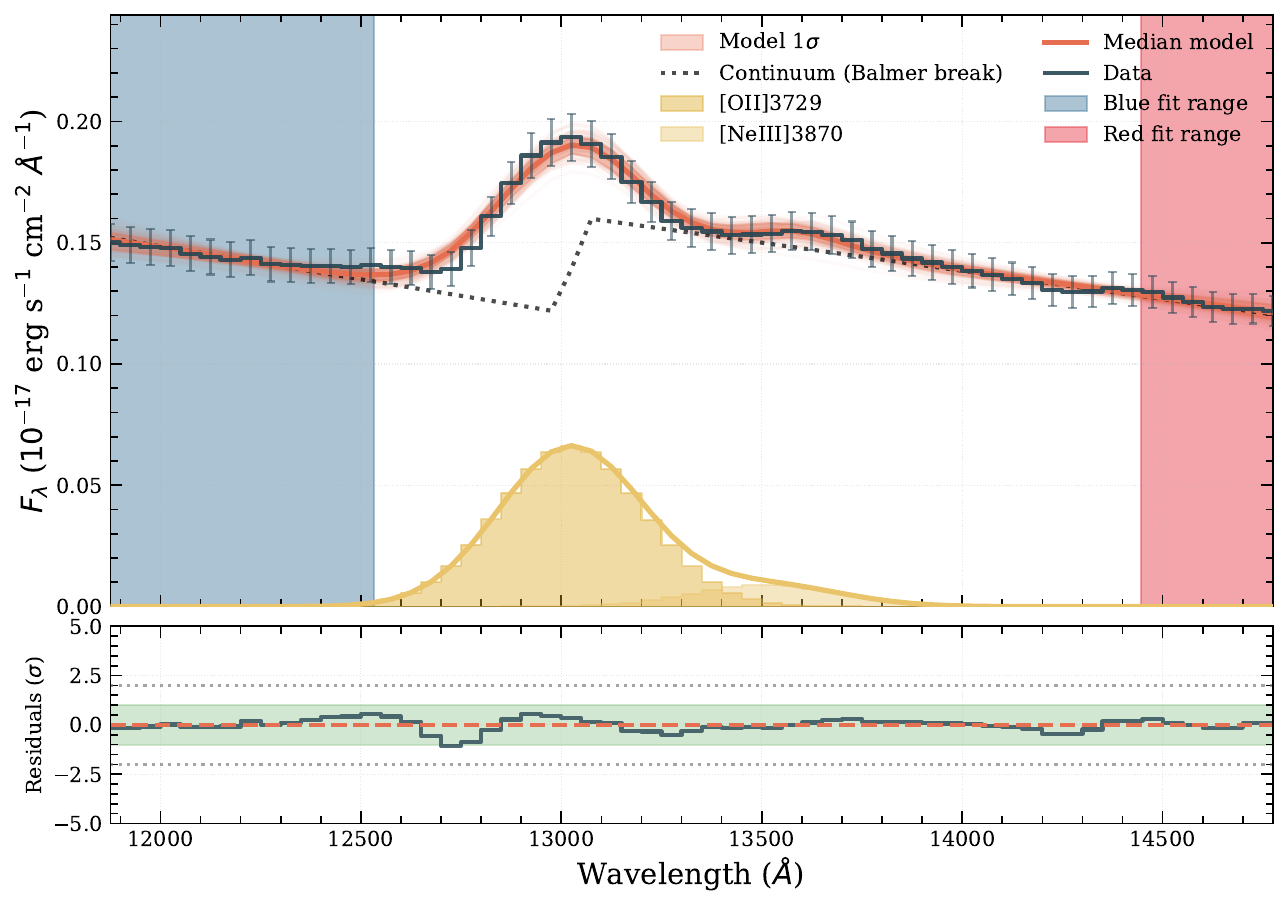} \\
    \includegraphics[width=0.99\linewidth]{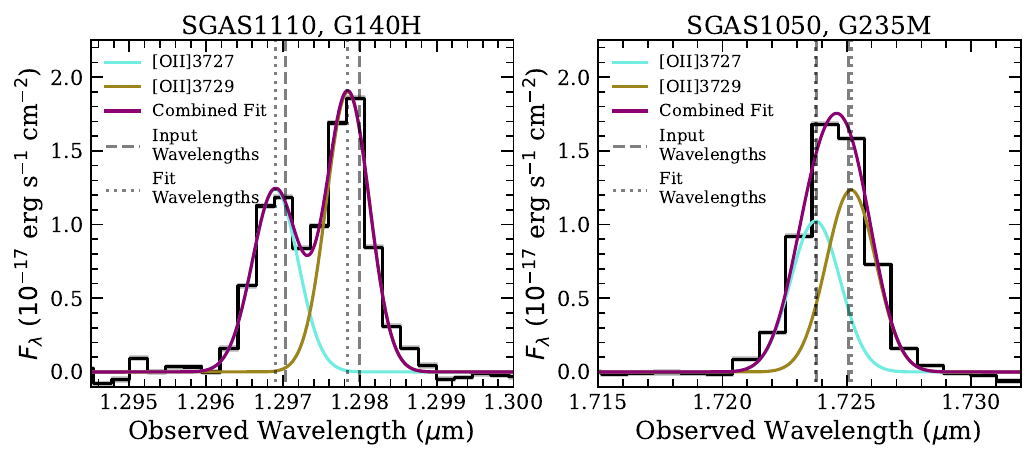}
    \caption{The top panel shows an example joint fit of the \oiibrightdbl\ doublet and the Balmer break for \elf. The joint \oii-Balmer break fits to the prism spectra yield \oiibrightdbl\ fluxes consistent with those measured in the high (medium) resolution spectrum for \elf\ (\fiddy). The \oii\ fits for the G140H and G235M spectra of \elf\ and \fiddy, respectively, are shown in the lower panels. In each panel, the dashed vertical lines show the initial line centers obtained by redshifting the vacuum wavelengths of the \oiibrightdbl\ lines, and the dotted vertical lines show the best-fit line centers, which are consistent with the redshifted vacuum wavelengths within one resolution element. The strong blending of the \oii\ lines in the G235M grating leads us to interpret the derived density from these lines with caution. }
    \label{fig:o2bb}
\end{figure}

We fit nearby and blended lines simultaneously. The medium-resolution data for \fiddy\ include two strongly blended sets of lines, namely the \oiibrightdbl\ doublet and \neiiib$+$\hepsilon. We fit these strongly blended pairs with two Gaussians each, with the line width fixed to the NIRSpec instrumental dispersion; however, we encourage caution when interpreting these reported individual line fluxes, as the relative contributions are difficult to constrain with the medium-resolution data (see Figure \ref{fig:o2bb}). 

The G235M spectrum for \fiddy\ includes a detection of the \arivdbl\ doublet, of which the bluer \ariva\ line is strongly blended with the neighboring \heibeiargon\ line. We fit the a single Gaussian to the \ariva\ and \heibeiargon\ lines, and then calculate the expected contribution of the \heibeiargon\ line based on the flux of the nearby \heia\ line assuming a temperature of $1.38\times10^4$ K, and a density of $400 \text{ cm}^{-3}$, similar to previous work \citep{welch_sunburst_2025}. The calculated flux of \heibeiargon\ and the corrected flux of \ariva\ are reported in Table \ref{table:flux}.

The low resolution of the prism data also causes many lines to be strongly blended. We fit \oiiauroraldbl\ in the prism spectra as a single Gaussian because the line separations are well below the resolution of the prism. We also fit the \siiib$+$Paschen-$\epsilon$ lines as a single Gaussian in the prism spectra. The \siiib\ fluxes reported in Table \ref{table:flux} do not include any corrections for contamination by Pa-$\epsilon$. Based on measured fluxes of the redder Paschen series lines, we expect this correction to be significant for \fiddy\ and \batleth, and insignificant for all other targets. %\red{NOTE - maybe do a quick correction for these?}

The \oiibrightdbl\ lines are covered in the grating data for \fiddy\ and \elf. 
For the other targets, we measure \oiibrightdbl\ flux from the prism data. 
In the prism spectra, \oiibrightdbl\ is strongly blended with the Balmer break, making our standard continuum subtraction method inaccurate. 
We therefore employ a joint \oiibrightdbl\ and Balmer break fitting method. 
We measure the continuum on either side of the Balmer break using a linear fit in the range $3400\leq \lambda \leq 3600$ \AA\ blueward of the break, and $4150\leq\lambda\leq 4250$ \AA\ redward of the break, following previous works \citep[][and references therein]{wang_rubies_2024,mintz_taking_2026,khullar_caught_2026}. 
The \oiibrightdbl\ and blended \neiiia\ and HeI$\lambda3889$ emission lines are included in the fitting process as single Gaussian lines with widths fixed based on the NIRSpec dispersion files, as with our standard line fitting described above. 
The relative redshifts of the Balmer break and emission lines are allowed to vary within 0.02 of the systemic redshift to account for any possible offsets between $z_{\text{stars}}$ and $z_{\text{gas}}$. 

An example \oiibrightdbl-plus-Balmer break fit for \elf\ is shown in Figure \ref{fig:o2bb}.
We find that the \oiibrightdbl\ fluxes measured with the prism fitting method for \elf\ and \fiddy\ match the fluxes from the grating measurements within uncertainties, indicating that this method is accurately recovering the \oiibrightdbl\ flux. 
The \oiibrightdbl\ fluxes are reported in Table \ref{tab:fluxes}. 
Though we fit the blend of \neiiia\ and HeI$\lambda3889$ in this process, we choose not to report the flux of this blend due to the low SNR of the line in the prism data. 

Finally, we note that the prism spectrum for \fiddy\ contains several possible rest-UV emission lines, though the low resolution and low sensitivity of the prism below 1$\mu$m observed wavelength makes it difficult to extract meaningful information from these features. We do fit the heavily blended \siliiidbl\ and \ciiidbl\ lines with a single Gaussian for this source, and the measured flux is reported in Table \ref{tab:fluxes}. However, we do not use this measurement for the present analysis, as the doublets are too heavily blended to provide either density constraints or reliable C or Si abundance constraints. We additionally fit the \niiiuv\ complex and the blend of \oiiiuvdbl\ and \heiiuv\ with single Gaussian profiles. Because the \oiiiuvdbl\ and \heiiuv\ lines are heavily blended at the prism wavelength, we do not attempt to use this fit for additional constraints on the O$^{++}$ electron temperature, and instead only report the temperature based on the \oiiiauroral\ flux. 

\subsection{Reddening Corrections}
We correct our measured emission line fluxes for Milky Way dust attenuation using the dust law of \cite{cardelli_relationship_1989}, with $E(B-V)$ values queried from the dust map of \cite{schlafly_measuring_2011} accessed via the NASA/IPAC Infrared Science Archive\footnote{\href{https://irsa.ipac.caltech.edu/applications/DUST/}{https://irsa.ipac.caltech.edu/applications/DUST/}}.

We correct for dust attenuation within our target galaxies using the Balmer decrement \halpha/\hbeta.
We calculate the expected Balmer decrement \halpha/\hbeta\ assuming an initial temperature of $10^4$K and initial density of $100\text{cm}^{-3}$, and compare the theoretical Balmer decrement to our measured \halpha/\hbeta\ ratio to obtain an initial $E(B-V)$.
Here, we assume the dust law of \cite{calzetti_dust_2000}, which was calibrated on local starburst galaxies likely to be representative of our star-forming galaxies at Cosmic Noon.
We calculate the uncertainty in the $E(B-V)$ value by randomly sampling the \halpha\ and \hbeta\ fluxes assuming a Gaussian distribution with standard deviation equal to the line flux uncertainty, then calculating $E(B-V)$ for each of 100 samples.
The resulting standard deviation is what we quote as the uncertainty in $E(B-V)$. 
We report the values and uncertainties for each target in Table \ref{table:flux}. 

The shape of the nebular attenuation curve may differ for high-$z$ galaxies \citep[e.g.,][]{sanders_aurora_2025, reddy_aurora_2026}. 
While multiple Balmer and Paschen lines are available for our sample, we leave detailed analysis of the nebular attenuation curves of these galaxies for future work.

\section{Star Formation Histories \& SED Modeling} \label{sec:sedfits}

\begin{table*}
    \centering
    %\scriptsize
    \caption{LEGGOS Sample Properties}
    \label{tab:sample}
    \begin{tabular*}{0.95\linewidth}{cccccccc}
    \hline \hline \\[-1.5ex]
    Target   & $z_{spec}$ & $\mu$      & $\log(M_*/M_{\odot})$ & SFR (SED)                    & sSFR                     & Age (UV)  & Age$_{MW}$ (Opt.) \\
             &            &            &                       & ($M_{\odot}\text{ yr}^{-1}$) & (yr$^{-1}$)              & (Myr)     & (Myr)              \\
    \hline  \\[-1.5ex]
    \fiddy   & 3.625      & $49\pm5$   & $8.08 \pm 0.04$       & $12.5^{+1.6}_{-1.2}$         & $\leq-7.0$               & --        & $\leq 10$         \\
    \eye     & 3.074      & $6\pm 1$   & $11.1 \pm 0.1$        & $62^{+14}_{-9}$              & $-9.31^{+0.04}_{-0.05}$  & $29\pm 1$ & $1460^{+50}_{-70}$ \\
    \batleth & 2.825      & $35\pm9$   & $9.01\pm0.15$         & $14^{+5}_{-3}$               & $-7.86^{+0.06}_{-0.12}$  & $15\pm 3$ & $40^{+20}_{-10}$    \\
    \geese   & 2.762      & $75\pm30$  & $8.4 \pm 0.2$         & $1.3^{+1.0}_{-0.4}$          & $-8.29^{+0.25}_{-0.06}$  & $21\pm 3$ & $1180^{+120}_{-200}$ \\
    \elf     & 2.481      & $10\pm2$   & $9.2 \pm 0.1$         & $8.2^{+2.0}_{-1.4}$          & $-8.29^{+0.06}_{-0.04}$  & --        & $130^{+10}_{-20}$     \\
    \\[-1.5ex]
    \hline
    \end{tabular*}
    \tablefoot{ Properties of the galaxies analyzed in this work. Magnifications ($\mu$) are reported for the apertures used for spectral/photometric extraction. Stellar mass, SFR, sSFR, and mass-weighted ages (Age$_{MW}$) are derived from \texttt{prospector} SED fitting of the JWST rest-optical imaging and spectroscopy, described in Section \ref{sec:sedfits}. Ages derived from rest-UV stellar population fits from \cite{chisholm_constraining_2019} are reported for comparison, using their \texttt{Starburst99} model fits. }

\end{table*}

\begin{figure}
    \centering
    \includegraphics[width=\linewidth]{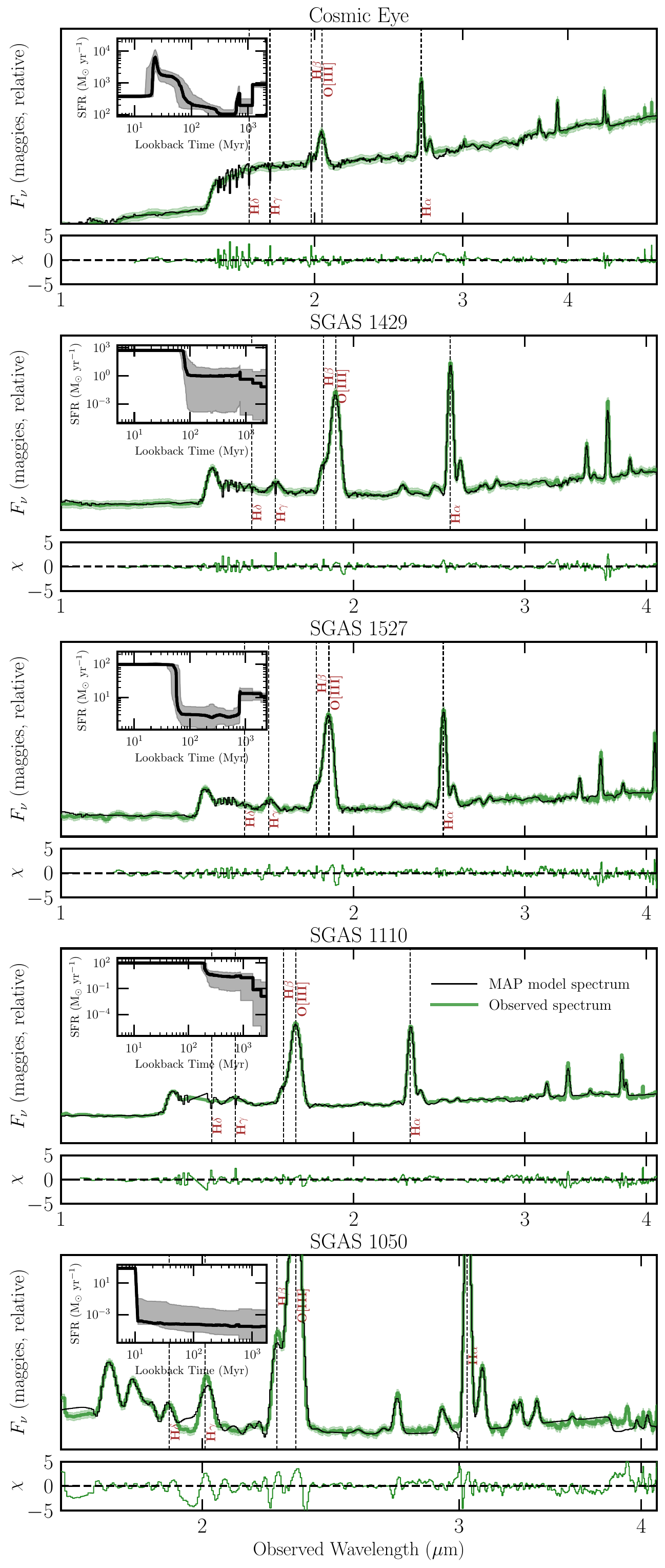} \\
    \caption{SED fits for all targets in our sample are shown. The black line is the best-fit SED model from \texttt{prospector}, while the green line and shaded region are the spectrum and $1\sigma$ uncertainty, respectively. The SFH for each galaxy is shown in black in the inset panel, with the grey region representing the $1\sigma$ uncertainty on the SFH. Every galaxy in our sample has undergone significant star formation within the last 100 Myr.  }
    \label{fig:sedfits1}
\end{figure}

We infer star formation histories (SFHs) for our five targets using stellar population synthesis (SPS) modeling via Bayesian SED fitting. We follow the modeling methodology described in \cite{khullar_leggos_2026}, and we summarize the key points here. 

We utilize both NIRCam photometry and NIRSpec spectroscopy in fitting the SEDs of these galaxies. We measure NIRCam photometry in an aperture defined by the SNR(\halpha)$>5$ spaxels measured in the NIRSpec IFU data, as described in Section \ref{subsec:ext1d}. In cases where a foreground lens galaxy is present in the vicinity of our target lensed arcs, we remove the light of the foreground galaxy by modeling it with a 2D Sersic profile, then subtracting that model from the image prior to calculating the aperture photometry. After the foreground galaxy contribution is subtracted, we calculate the aperture photometry using \texttt{photutils} \citep{bradley_astropyphotutils_2025}. 

We fit the NIRCam photometry and NIRSpec spectroscopy simultaneously using the SED fitting code \texttt{prospector} \citep{johnson_bd-jprospector_2017, leja_deriving_2017,johnson_stellar_2021}. \texttt{Prospector} utilizes Flexible Stellar Population Synthesis (FSPS) stellar population synthesis models \citep{conroy_propagation_2009, conroy_propagation_2010}, the MILES spectral library \citep{sanchez-blazquez_medium-resolution_2006,falcon-barroso_updated_2011}, and the MIST isochrones \citep{choi_mesa_2016, dotter_mesa_2016}. We implement a non-parametric SFH model with flexible age bins as described in \cite{khullar_leggos_2026}. This includes a fixed youngest age bin size of 10 Myr. We assume a \cite{chabrier_galactic_2003} initial mass function and a \cite{kriek_dust_2013} dust law with $A_v$ and dust index as free parameters, with doubled attenuation around young ($<10^7$ yr old) stars \citep{wild_star_2020,suess_rest-frame_2022,setton_desi_2023}. We marginalize over nebular line emission using Gaussian emission line profiles, meaning that the emission line strengths do not directly affect the inferred SFH. We sample the posterior distribution using the \texttt{dynesty} dynamic nested sampling package \citep{speagle_dynesty_2020}.

Table \ref{tab:sample} presents the parameters inferred from SED fitting, and Figure \ref{fig:sedfits1} displays the best-fit models and SFHs. The apparent discontinuity in the plotted spectrum of \elf\ is due to masking a detector artifact. The affected region is not used in the SED fitting process. Inferred stellar masses and star formation rates (SFRs) of gravitationally lensed galaxies are directly proportional to the magnification. Table \ref{tab:sample} reports the magnification-corrected values of both $M_*$ and SFR, along with the magnification $\mu$ used for this correction. Magnification uncertainties are included in the quoted $M_*$ and SFR values by drawing 300 random samples from the distribution of magnifications and magnified parameters $\mu M_*$ and $\mu\text{SFR}$, assuming each distribution is a Gaussian with mean and standard deviation given by the best-fit value and uncertainties, respectively. The final uncertainties quoted in Table \ref{tab:sample} are the 16th and 84th percentiles of the resulting demagnified parameter distributions.

Specific star formation rates (sSFR) are not affected by magnification, as the magnification dependence cancels out when dividing SFR by $M_*$. Stellar population ages are driven by the SED shape, and are therefore not affected by achromatic lensing magnification. For comparison, we also present the stellar population ages for \eye, \batleth, and \geese\ derived in \cite{chisholm_constraining_2019}, which are based on modeling of the rest-UV spectrum using \texttt{Starburst99} \citep{leitherer_starburst99_1999,leitherer_effects_2014}. The rest-UV stellar population ages tend to be lower than the rest-optical age estimates, since the rest-UV spectrum is primarily driven by recently formed massive stars, while the optical includes greater contributions from older stars. However, each stellar population age estimate is dependent on the SFH, and the uncertainties reported in Table \ref{tab:sample} are only statistical uncertainties, not accounting for systematics such as SFH assumptions. The SPS modeling in the rest-optical tends to struggle with stellar population ages less than $\sim 10$ Myr, and the youngest age bin in our analysis has been fixed to 10 Myr. The reported age for \fiddy\ is treated as an upper limit, as all of the star formation in our model has occurred within the last 10 Myr prior to observation.

%%%%%%%%%%%%%%%%%%%%%%%%%%%%%%%%%%%%%%%%%%%%%%%%%%%%%%%%%%%%%%
\section{Direct Temperatures, Densities, and Chemical Abundances} \label{sec:physconditions}

With the detection of faint temperature-sensitive emission lines, we characterize the physical conditions and chemical abundance patterns of the ISM in our galaxy sample. We use \texttt{PyNeb} \citep{luridiana_pyneb_2015} to calculate the electron densities, temperatures, and ionic abundances of each galaxy as described below. We adopt a three-zone model of the ISM for these calculations, described by ions with similar ionization potentials (I.P.). The low-ionization zone includes the ions  N$^+$ (I.P. 14.53 eV), O$^+$ (I.P. 13.62 eV), and S$^+$ (I.P. 10.36 eV), the intermediate zone includes S$^{+2}$ (I.P. 23.34 eV) and Ar$^{+2}$ (I.P. 27.63 eV), and the high-ionization zone includes O$^{+2}$ (I.P. 35.12 eV), Ne$^{+2}$ and Ar$^+{3}$ (I.P. 47.40 eV). 

%We use an iterative approach to jointly constrain the temperature and density of each target, starting with a fiducial temperature of $10^4$ K and a fiducial density of 100 cm$^{-3}$, and updating the density and temperature based on available line ratios until we reach convergence. We set convergence to be an average change in temperature and density values less than 15, which is less than the statistical uncertainties for each measurement. We then use the most recent low ionization zone temperature to calculate the final density, choosing to calculate a final density first because the density from \sii\ has a weak dependence on temperature. We then use our final density estimate to calculate temperatures for each ionization zone. Uncertainties on both quantities are estimated using 300 random samples of the relevant line fluxes, assuming the line flux is gaussian distributed with mean and sigma given by the best-fit and uncertainty in the line flux measurements. All temperatures and densities are calculated using the \texttt{getTemDen} function in pyneb. 

\begin{table}
    \centering
    \scriptsize
    \caption{Atomic Data Sources}
    \label{tab:atomicdata}
    \begin{tabular*}{0.9\linewidth}{l l l}
    \hline \hline
        Ion & Transition Probabilities & Collision Strengths \\
    \hline 
        N$^+$ & \cite{froese_fischer_breit-pauli_2004} & \cite{tayal_electron_2011} \\
        O$^+$ & \cite{froese_fischer_breit-pauli_2004} & \cite{kisielius_electron-impact_2009} \\
        O$^{++}$ & \cite{froese_fischer_breit-pauli_2004} & \cite{storey_collision_2014} \\
        Ne$^{++}$ & \cite{froese_fischer_breit-pauli_2004} & \cite{mclaughlin_large-scale_2011} \\
        S$^+$ & \cite{rynkun_theoretical_2019} & \cite{tayal_breit-pauli_2010} \\
        S$^{++}$ & \cite{tayal_collision_2019} & \cite{grieve_electron-impact_2014} \\
        Ar$^{++}$ & \cite{mendoza_transition_1983} & \cite{munoz_burgos_electron-impact_2009} \\
        Ar$^{+3}$ & \cite{rynkun_theoretical_2019} & \cite{ramsbottom_effective_1997} \\
    \hline
    \end{tabular*}

\end{table}

\subsection{Densities} \label{subsec:dens}

The NIRSpec data cover two density-sensitive doublets, \siidbl\ and \oiibrightdbl. The \siidbl\ doublet falls in the medium-resolution G235M spectra for each target, so we adopt the \siidbl\ density as our fiducial density estimate to ensure consistency across the sample. 

Density estimates from \sii\ and \oii\ are weakly sensitive to temperature. We iteratively calculate the density and temperature using all available constraints for each target until the estimates converge, where convergence is defined as the average of the differences in temperature and density between the previous iteration and the current iteration being less than 15 (i.e. less than the statistical uncertainties on any of our measured temperatures or densities). We then use the most recent low-ionization temperature to calculate final density values. Table \ref{tab:results} reports the density estimates for each galaxy. Uncertainties are estimated by drawing 300 random samples of the relevant emission line fluxes, assuming a Gaussian distribution with a mean and standard deviation defined by the line flux and uncertainty, respectively. %Resulting density distributions are approximately Gaussian and centered on the best-fit density value for all measurements except the \ariv\ density in \fiddy, where larger emission line flux uncertainties result in a skewed density distribution. We report the 16th and 84th percentile uncertainties in Table \ref{tab:results} for the \ariv\ density.

Although the data cover the \oiibrightdbl\ doublet for all targets, only two targets (\elf\ and \fiddy) observe these lines in a grating with sufficiently high resolution to attempt a density constraint. The other targets cover the \oiibrightdbl\ lines in the prism spectra, which have too low spectral resolution to resolve the doublet. \fiddy\ covers the \oiibrightdbl\ lines in the G235M grating, which is not high enough spectral resolution to fully resolve the doublet, though we do see that the blended lines appear broader than any of the nearby unblended emission lines. We therefore fit these as two Gaussians, as described in Section \ref{subsec:linefits}. The resulting density constraint is consistent with the density from \siidbl, though we advise readers to interpret this result with caution since the \oiibrightdbl\ lines are still heavily blended in the G235M data. 

\elf\ includes G140H observations which are high enough resolution to resolve the \oiibrightdbl\ doublet. We find that both the \oiibrightdbl\ and \siidbl\ doublet ratios are slightly above the theoretical maximum, though they are both consistent with the maximum within $1\sigma$ uncertainties. We adopt a density of $100 \pm 50\text{ cm}^{-3}$ for \elf.

Recent work \citep[e.g.,][]{martinez_under_2025,arellano-cordova_self-consistent_2026} has highlighted the importance of density stratification in nebulae, as the low-ionization tracers such as \oii\ (I.P. 13.62 eV) and \sii\ (I.P. 10.36 eV) can underestimate the average gas density as the average density increases, thereby biasing abundance measurements. We therefore explore additional density tracers at higher ionization states for our targets. The NIRSpec data cover the \arivdbl\ density-sensitive doublet (I.P. 47.40 eV), which can be a good probe of high-ionization densities. We only detect this doublet in one target (\fiddy). The density measurement is complicated by the blending of the \ariva\ and \heibeiargon\ lines. While we attempt to subtract out the \heibeiargon\ flux to recover the \ariva\ line (see Sec. \ref{subsec:linefits}), we note that this process adds substantial uncertainty to the resulting density estimate. We find that \arivdbl\ density is consistent within $1\sigma$ uncertainty of the low-density limit of $1000 \text{ cm}^{-3}$, suggesting that we do not see significant contribution from high-density regions within this galaxy. 

The near-UV \ciiidbl\ or \siliiidbl\ doublets offer additional useful density constraints, but unfortunately these lines are only covered in the prism observations where the spectral resolution is too low to resolve the doublet. We therefore instead use the available ground-based spectra to attempt to constrain the \ciiidbl\ density for our sample. 

\fiddy\ has published densities from both \ciiidbl\ and \siliiidbl; both are found to be in the low-density limit for these ratios, resulting in a density constraint of $\leq 10^3 \text{ cm}^{-3}$ \citep{bayliss_physical_2014}. These results support our use of the low-ionization density as our primary constraint for \fiddy, as one would expect these intermediate- to high-ionization indicators to yield higher densities if the galaxy were dominated by high-density high-ionization gas. 

\elf\ has a detection of \ciiidbl\ \citep{johnson_star_2017}, however, the doublet appears as a single line due to a combination of low SNR and spectral resolution. This arc also has higher spectral resolution observations from MMT Blue Channel data (as described in Sec. \ref{subsec:grounddata}), however, the \ciiidbl\ lines are not significantly detected in this spectrum. We cannot robustly constrain the higher-ionization density for \elf\ with existing data. 

\geese\ has weak detections of the \ciiidbl\ doublet in the stacked Magellan/MagE spectrum described in Section \ref{subsec:grounddata} from the MEGaSaURA program \citep{rigby_magellan_2018}. We fit the doublet using the same procedure detailed in Section \ref{subsec:linefits}, but without fixing the linewidths. We find fluxes of $F(1907) = 1.8\pm0.3 \times10^{-17} \text{ erg s}^{-1}\text{ cm}^{-2}$ and $F(1909) = 0.9\pm0.3 \times10^{-17} \text{ erg s}^{-1}\text{ cm}^{-2}$, resulting in a best-fit ratio of $F(1907)/F(1909) = 1.9 \pm 0.3$. This ratio is greater than the theoretical upper limit for the doublet, but the measured ratio is consistent with the low-density limit within $1\sigma$ uncertainties. We adopt an upper limit on the density from \ciiidbl\ of $n_e < 10^3 \text{ cm}^{-3}$, which increases our confidence in using the low-ionization \siidbl\ density as our fiducial estimate. 

\batleth\ and \eye\ do not have significant \ciiidbl\ detections in existing ground-based data, so we do not have any higher-ionization density constraints for these objects. Deeper rest-UV spectra could improve the characterization of the physical conditions within each of these arcs.

\begin{table*}[]
    \centering
    \caption{Nebular Temperatures, Densities, and Abundances}
    \begin{tabular}{c c c c c c}
    \hline \hline \\
         & SGAS1050 & Cosmic Eye & SGAS1429 & SGAS1527 & SGAS1110  \\
    \hline \\
$z$ & $3.625$ & $3.074$ & $2.825$ & $2.762$ & $2.481$  \\ 
\hline \\[-1.5ex]
$n_e$[SII] (cm$^{-3}$) & $400 \pm 70$ & $490 \pm 30$ & $280 \pm 30$ & $170 \pm 80$ & $100 \pm 50$\tablefootmark{b}  \\ 
$n_e$[OII] (cm$^{-3}$) & $360 \pm 100$\tablefootmark{a} & ... & ... & ... & $100 \pm 50$\tablefootmark{b}  \\ 
$n_e$[ArIV] (cm$^{-3}$) & $2800^{+6700}_{-1900}$ & ... & ... & ... & ...  \\ [1.2ex]
\hline \\[-1.5ex]
$T_e$[OII] (K) & $14500 \pm 700$ & $8100 \pm 800$ & $9100 \pm 500$ & $11000 \pm 1200$ & $12000 \pm 1100$  \\ 
$T_e$[NII] (K) & $36000 \pm 10600$ & ... & ... & ... & ...  \\ 
$T_e$[SIII] (K) & $10900 \pm 700$ & ... & $9800 \pm 700$ & ... & $15500 \pm 2000$  \\ 
$T_e$[OIII] (K) & $13800 \pm 100$ & ... & $11000 \pm 600$ & ... & ...  \\ 
$T_e$(Low) (K) & $14500 \pm 700$ & $8100 \pm 800$ & $9100 \pm 500$ & $11000 \pm 1200$ & $12000 \pm 1100$  \\ 
$T_e$(Int.) (K) & $10900 \pm 700$ & $7800 \pm 1300$ & $9800 \pm 700$ & $11900 \pm 1900$ & $15500 \pm 2000$  \\ 
$T_e$(High) (K) & $13800 \pm 100$ & $8500 \pm 1100$ & $11000 \pm 600$ & $11100 \pm 1400$ & $13400 \pm 1500$  \\ 
\hline \\[-1.5ex]
O$^+$/H$^+$ ($\times 10^5$) & $1.2 \pm 0.2$ & $44 \pm 23$ & $15 \pm 4$ & $6 \pm 3$ & $3.8 \pm 1.3$  \\ 
O$^{++}$/H$^+$ ($\times 10^5$) & $9.8 \pm 0.2$ & $18 \pm 10$ & $10.7 \pm 1.9$ & $13 \pm 6$ & $8 \pm 2$  \\ 
12+log(O/H) & $8.04 \pm 0.01$ & $8.79 \pm 0.18$ & $8.41 \pm 0.07$ & $8.28 \pm 0.15$ & $8.08 \pm 0.10$  \\ 
\hline \\[-1.5ex]
N$^+$/H$^+$ ($\times 10^7$) & $7.7 \pm 0.8$ & $288 \pm 85$ & $63 \pm 10$ & $18 \pm 5$ & $14 \pm 3$  \\ 
N$^++$/H$^+$ ($\times 10^5$) & $1.8 \pm 0.9$ & ... & ... & ... & ... \\
12+log(N/H) & $6.84 \pm 0.04$ & $7.6 \pm 0.2$ & $7.03 \pm 0.11$ & $6.76 \pm 0.15$ & $6.65 \pm 0.10$  \\ 
log(N/O) & $-1.20 \pm 0.08$ & $-1.19 \pm 0.26$ & $-1.39 \pm 0.13$ & $-1.52 \pm 0.24$ & $-1.43 \pm 0.17$  \\ 
\hline \\[-1.5ex]
Ne$^{++}$/H$^+$ ($\times 10^5$) & $2.23 \pm 0.06$ & ... & ... & ... & $2.3 \pm 0.8$  \\ 
ICF(Ne) & $1.02 \pm 0.10$ & ... & ... & ... & $1.29 \pm 0.13$  \\ 
12+log(Ne/H) & $7.36 \pm 0.05$ & ... & ... & ... & $7.47 \pm 0.17$  \\ 
log(Ne/O) & $-0.68 \pm 0.05$ & ... & ... & ... & $-0.61 \pm 0.19$  \\ 
\hline \\[-1.5ex]
S$^+$/H$^+$ ($\times 10^7$) & $1.73 \pm 0.18$ & $22 \pm 7$ & $11.3 \pm 1.8$ & $5.3 \pm 1.4$ & $3.9 \pm 0.7$  \\ 
S$^{++}$/H$^+$ ($\times 10^7$) & $17 \pm 2$ & $60 \pm 27$ & $31 \pm 4$ & $13 \pm 4$ & $9.3 \pm 1.7$  \\ 
ICF(S) & $1.87 \pm 0.19$ & $0.95 \pm 0.10$ & $0.97 \pm 0.10$ & $1.15 \pm 0.11$ & $1.13 \pm 0.11$  \\ 
12+log(S/H) & $6.55 \pm 0.07$ & $6.89 \pm 0.16$ & $6.62 \pm 0.06$ & $6.32 \pm 0.11$ & $6.17 \pm 0.07$  \\ 
log(S/O) & $-1.49 \pm 0.07$ & $-1.90 \pm 0.24$ & $-1.80 \pm 0.10$ & $-1.95 \pm 0.18$ & $-1.90 \pm 0.12$  \\ 
\hline \\[-1.5ex]
Ar$^{++}$/H$^+$ ($\times 10^7$) & $2.23 \pm 0.06$ & $5 \pm 2$ & $4.0 \pm 0.5$ & $3.0 \pm 0.9$ & $1.7 \pm 0.4$  \\ 
Ar$^{+3}$/H$^+$ ($\times 10^7$) & $0.9 \pm 0.3$ & ... & ... & ... & ...  \\ 
ICF(Ar) & $1.00 \pm 0.10$ & $1.19 \pm 0.12$ & $1.14 \pm 0.11$ & $1.06 \pm 0.11$ & $1.06 \pm 0.11$  \\ 
12+log(Ar/H) & $5.50 \pm 0.13$ & $5.78 \pm 0.17$ & $5.66 \pm 0.07$ & $5.50 \pm 0.13$ & $5.25 \pm 0.11$  \\ 
log(Ar/O) & $-2.55 \pm 0.13$ & $-3.01 \pm 0.25$ & $-2.75 \pm 0.10$ & $-2.78 \pm 0.20$ & $-2.82 \pm 0.15$  \\ 

        \hline
    \end{tabular}
    \tablefoot{Measured physical conditions and chemical abundances of the LEGGOS galaxies. ICFs reported in this table are calculated using the relations from \cite{izotov_chemical_2006}. \\
    \tablefoottext{a}{Measured from the medium-resolution G235M spectrum. We suggest interpreting this density result with caution, as the lines remain strongly blended.} \\
    \tablefoottext{b}{Best-fit line ratios are outside the theoretical range, but consistent within $1\sigma$ uncertainties with the low-density limit. We therefore adopt $n_e=100\pm 50\text{ cm}^{-3}$ as a fiducial density for this target. }
    }
    \label{tab:results}
\end{table*}

\subsection{Temperatures} \label{subsec:temps}

The strength of auroral lines of a particular ion depends on the temperature of the gas which contains that ion. It is possible that ions with different ionization potentials occupy different regions of the galaxy and may have different temperatures. Ideally, temperature diagnostics in all three ionization zones ($T_e(\text{High})$, $T_e(\text{Int.})$, $T_e(\text{Low})$) can be measured to fully characterize the physical conditions of the gas within each galaxy. When some ionization zones do not have measured temperatures, it is possible to use temperature-temperature scaling relations to infer the missing temperatures \citep[e.g.,][]{garnett_electron_1992,rogers_chaos_2021}. 

We measure electron temperatures using the temperature-sensitive auroral lines \oiiiauroral, \niiauroral, \siiiauroral, and \oiiauroraldbl, spanning all three ionization zones. Only the low-ionization \oii\ temperature is measured across the full sample. For each galaxy, we calculate temperatures using the \texttt{getTemDen} function from \texttt{PyNeb}, using the \sii\ density as described in the previous section. We only include temperatures based on emission lines with SNR$>3$. \elf\ exhibits a weak emission feature at \oiiiauroral, however, we do not use this line to measure a temperature due to it being less than our $3\sigma$ threshold. Table \ref{tab:results} reports temperature measurements for each galaxy analyzed in this work. 

\fiddy\ is the only galaxy in our sample with all four temperatures measured. However, we find that the \nii\ temperature $T_e(\text{N}^+)\sim36,000$ K is unphysically high, and we conclude that this is driven by a spurious flux excess around the \niiauroral\ line; we do not use the \nii\ temperature for further analysis. 

\begin{figure}[h!]
    \centering
    \includegraphics[width=0.95\linewidth]{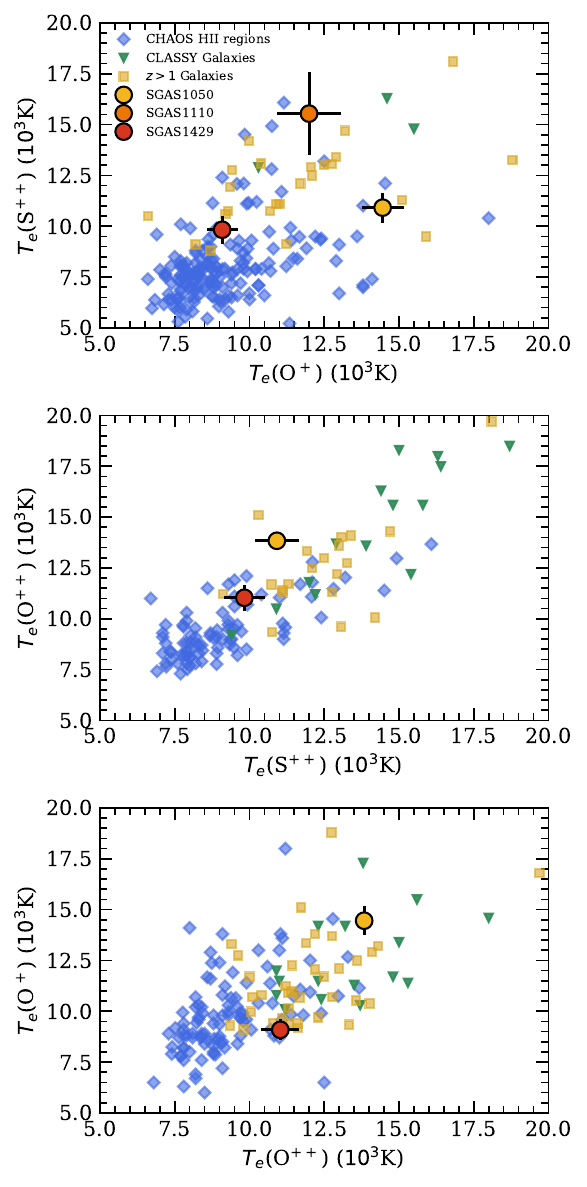}
    \caption{Temperature measurements for \fiddy, \elf, and \batleth, for which multiple temperatures are available. We plot these temperatures alongside comparison samples of local \hii\ regions from the CHAOS program, low-$z$ star-forming galaxies from the CLASSY program, and a compilation of high-$z$ objects, as described in Section \ref{sec:discussion}. Our galaxy sample generally matches temperature trends previously observed in both low- and high-$z$ galaxies and \hii\ regions. }
    %two other lensed objects with multiple temperatures measured -- SGAS1723 \citep{welch_templates_2024} and the Ly-C leaker in the Sunburst Arc \citep{welch_sunburst_2025}. We also show temperatures for $z=0$ HII regions from the CHAOS program \citep{berg_chaos_2015,croxall_chaos_2015,croxall_chaos_2016,berg_chaos_2020,rogers_chaos_2021}, and $z=0$ galaxies from the CLASSY program \citep{arellano-cordova_classy_2024}. We also show a sample of high-redshift objects with multiple temperature measurements, including from CECILIA \citep{rogers_cecilia_2026}, EXCELS \citep{arellano-cordova_jwst_2025,stanton_jwst_2025}, MARTA \citep{curti_marta_2025,cataldi_marta_2025}, and AURORA \citep{sanders_aurora_2025}. }
    \label{fig:temps}
\end{figure}

Three galaxies (\fiddy, \elf, and \batleth) have multiple temperature measurements, and two of these (\fiddy\ and \batleth) include temperatures in all three ionization zones. We plot these temperatures in Figure \ref{fig:temps}, alongside temperatures measured in other nearby \hii\ regions \citep{berg_chaos_2015,croxall_chaos_2015,croxall_chaos_2016,berg_chaos_2020,rogers_chaos_2021}, low-$z$ galaxies selected as high-$z$ analogs \citep{arellano-cordova_classy_2024}, and a compilation of other high-$z$ galaxies from the literature \citep{rogers_cecilia_2026,arellano-cordova_jwst_2025,stanton_jwst_2025,curti_marta_2025,cataldi_marta_2025,sanders_aurora_2025}. We find that our measured temperatures generally match the temperature-temperature scaling seen in both local and high-$z$ galaxies, though we note that the correlations involving the \oii\ temperature tend to show the most scatter \citep[e.g.,][]{rogers_chaos_2021}. Additionally, the number of high-$z$ galaxies with multiple temperatures are still somewhat limited owing to the faintness of the $T_e$-sensitive lines, though JWST observations have steadily increased the number of galaxies with such measurements in recent years. 

\geese\ does not have an \oiiiauroral\ detection because this line is off the blue end of the G235M spectrum at $z=2.762$, and no bluer grating observations were obtained. Based on the temperature measured from the \oii\ lines and the measured flux of \hbeta, we expect that a similarly deep observation of this galaxy with the G140M grating would have detected the \oiiiauroral\ line. While the line is covered in the prism, the low resolution means that it is too strongly blended with H$\gamma$ to robustly measure. 

When a particular ionization zone does not have a measured temperature, we use empirical temperature scaling relations to infer the $T_e$ of the relevant zone. We use the temperature relations of \cite{rogers_chaos_2021}, which were calibrated based on the CHAOS sample of \hii\ regions in five nearby spiral galaxies. The CHAOS sample spans a wide range of temperatures and oxygen abundances, and the large number of data points enables robust determinations of the correlations between temperatures in different ionization zones. Additionally, our galaxy sample matches the range of O/H abundances probed by the CHAOS sample, suggesting that this is a reasonable choice for our galaxies. While some higher-redshift temperature scaling relations exist \citep[e.g.,][]{cataldi_marta_2025}, these are thus far based on fewer data points and cover fewer temperatures than the locally calibrated versions, so we elect to use the local calibrations for this work. 

We prioritize using the $T_e\oiii - T_e\siii$ relation where possible. The \oiiauroraldbl\ lines, though bright, are the most susceptible to contamination from dielectronic recombination \citep{rubin_noncollisional_1986,liu_chemical_2001}, and have typically been deprioritized in abundance studies in local galaxies when the \niiauroral\ line is available \citep[e.g.,][]{berg_chaos_2020,rogers_chaos_2021}. We therefore prefer the more reliable $T_e\oiii - T_e\siii$ relation wherever possible. When the \oiiauroraldbl\ lines are the only temperature-sensitive lines detected, i.e. in \eye\ and \geese, we make the assumption that $T_e\oii \sim T_e\nii$, and use the \nii-based temperature scaling relations to calculate higher-ionization temperatures. We include the intrinsic scatter reported in \cite{rogers_chaos_2021} when propagating uncertainties to inferred temperatures. When using $T_e\oii$ to calculate higher-ionization temperatures, we include the scatter in the $T_e\nii - T_e\oii$ relation from \cite{rogers_chaos_2021} as an additional uncertainty. 

The \oiii\ temperature can be biased high if a low density is assumed when the actual gas density is higher \citep{martinez_under_2025}. While we found little evidence for increased densities in high-ionization zones in the previous section (Sect. \ref{subsec:dens}), we perform a test for \fiddy\ in which we use the higher density from the \arivdbl\ doublet to recalculate the \oiii\ temperature, following previous work \citep{rogers_lbt_2026}. We find that the resulting \oiii\ temperature is consistent with our fiducial value within $1\sigma$ uncertainties, as is the resulting O$^{++}$ abundance. Thus we conclude that there is no evidence for temperature or abundance biases due to density stratification in \fiddy. Although we do not have well-measured higher ionization densities for the other targets, we conclude that the potential temperature bias is likely similarly low based on the low \sii\ densities measured.

\subsection{Chemical Abundances} \label{subsec:abunds}

The gas-phase abundances of ions relative to ionized hydrogen are a function of the collisionally-excited emission line strengths and the emissivities of the respective atomic transitions, namely 
\begin{equation}
    \frac{N(X^i)}{N(H^+)} = \frac{I_{\lambda(i)}}{I_{\hbeta}} \frac{j_{\hbeta}}{j_{\lambda(i)}} .
\end{equation}
The emissivity coefficients $J_{\lambda(i)}$ are a function of both the temperature and density of the gas. We calculate ionic abundances using the \texttt{getIonAbundance} function of \texttt{PyNeb} \citep{luridiana_pyneb_2015}, using the relevant electron temperatures and densities calculated as described above and the atomic data tabulated in Table \ref{tab:atomicdata}. All of our ionic abundances are presented in Table \ref{tab:results}.

We calculate uncertainties on the ionic abundances by sampling the emission line ratio and temperature uncertainties. We assume that both the line ratios and temperatures are described by a Gaussian distribution with the mean and standard deviation given by the measured value and uncertainty. We sample each of these distributions 100 times and calculate the ionic abundance for each combination. We use the half difference between the 84th and 16th percentiles of the resulting distribution as our uncertainty on the ionic abundance. The resulting distributions are approximately Gaussian and centered on the best-fit values for all ionic abundances. While density uncertainties do contribute to the overall uncertainty budget for the ionic abundances, the fractional uncertainties contributed by the densities are small relative to the temperatures; thus, we do not include density uncertainties in this calculation.

We calculate the total oxygen abundance as O/H = O$^+$/H$^+$ + O$^{++}$/H$^+$. We do not include any correction for unobserved O$^{+3}$, as studies of nearby extreme emission line galaxies (EELGs) have found that this high ionization state contributes $\lesssim2$\% of the total oxygen abundance \citep{berg_characterizing_2021,rickards_vaught_interstellar_2025}. While we do detect \oi\ emission in most of our galaxy spectra, we assume that most of this emission originates outside the \hii\ regions, and that the total contribution of \oi\ to the O/H budget is minimal. 

%When calculating oxygen abundances, we use the $T_e$(High) to calculate the O$^{++}$ abundance, and we use $T_e$(Low) to calculate the O$^+$ abundance. 

Depending on the ionization state of the gas, the nitrogen in nebulae can be contained in a combination of the N$^+$, N$^{++}$, and N$^{+3}$ states. However, only the N$^+$ ion is easily observable across the full galaxy sample. We therefore account for the missing higher-ionization nitrogen using the assumption that N/O $\simeq$ N$^+$/O$^+$ based on the similarity of the ionization potentials for the N$^+$ and O$^+$ ions \citep[14.53 eV and 13.62 eV, respectively;][]{peimbert_temperature_1967}. This assumption is commonly used at both low- and high-redshift \citep[e.g.,][]{berg_chaos_2020,rogers_chaos_2021,welch_templates_2024,arellano-cordova_jwst_2025,rogers_cecilia_2026}, and it has been found be valid to within $\sim10$\% in multiple studies \citep[e.g.,][]{nava_determination_2006,amayo_ionization_2021,martinez_under_2025}.

In \fiddy, the higher-ionization \niiiuv\ line is detected in the prism spectrum. We calculate the N$^{++}$/H$^+$ abundance using the strength of the \niiiuv\ line, assuming the high-ionization temperature derived from \oiii, and report the ionic abundance in Table \ref{tab:results}. We use the combined relative abundance $(\text{N}^+ +\text{N}^{++}) / (\text{O}^+ + \text{O}^{++})$ with the ionization corrections presented in \cite{martinez_under_2025}, linearly interpolating between their $n_e=10^2\text{ cm}^{-3}$ and $n_e=10^3\text{ cm}^{-3}$ fits, to derive the total $\log(\text{N/O}) = -0.8 \pm 0.2$. This value is $\sim 2\sigma$ higher than the value of $\log(\text{N/O})$ based on only the low-ionization ionic abundances, but we interpret this higher N/O value with caution due to the sensitivity of the rest-UV lines to dust corrections and to density and ionization variations \citep{martinez_under_2025}. We also note that the $\log(\text{N}^{++}/\text{O}^{++}) = -1.2\pm0.3$ derived based on a $2\sigma$ \niiiuv\ detection in higher spectral resolution GMOS data \citep{bayliss_physical_2014} matches our N$^+$/O$^+$ abundance, casting further doubt on the higher N/O from the prism \niiiuv\ detection. Deeper high spectral resolution rest-UV observations of \fiddy\ could better measure the N$^{++}$ abundance alongside other rest-UV physical condition tracers to provide a better constraint on the higher-ionization nitrogen abundance.

For Ne, S, and Ar, we correct for unobserved ionization states using the ionization correction factors (ICFs) calculated in \cite{izotov_chemical_2006}, which are calibrated based on photoionization models and include dependencies on both ionization conditions (parametrized by O$^+$/(O$^+$+O$^{++}$) and metallicity ($12+\log(\text{O/H})$). \cite{arellano-cordova_classy_2024} perform extensive tests of various ICFs using the CLASSY sample of local star-forming galaxies \citep{berg_cos_2022,james_classy_2022}, and find that the \cite{izotov_chemical_2006} consistently perform well. The CLASSY galaxies are considered analogues of high-$z$ galaxies, and the sample shows similar properties to the LEGGOS sample in $M_*$, SFR, and $12+\log(\text{O/H})$, suggesting that the CLASSY recommendations are applicable here. As a consistency check, we computed ICFs for each galaxy using the prescription of \cite{amayo_ionization_2021}, and found that in almost every case the resulting abundances of Ne, S, and Ar were consistent within uncertainties. The one exception (\fiddy) is discussed below. 

Following the recommendation of \cite{izotov_chemical_2006}, we linearly interpolate between the low, intermediate, and high metallicity ICFs to obtain the final correction based on the metallicity of each target. We assume a 10\% uncertainty in an attempt to account for the systematic uncertainty in the choice of ionization correction, following previous works \citep[e.g.,][]{rogers_chaos_2021,rogers_cecilia_2026}.

We only calculate neon abundances for targets where the \neiii\ lines are detected in one of the grating observations, as both \neiii\ lines are heavily blended with neighboring \hei\ and \hi\ recombination lines in the prism data.

\fiddy\ has a detection of the high-ionization \arivdbl\ lines, which we use to calculate the Ar$^{+3}$/H$^+$ abundance. For this target, we use the ICF(Ar$^{+2}$+Ar$^{+3}$) provided in Equation 23 of \cite{izotov_chemical_2006} rather than the ICF(Ar$^{+2}$) used for the other galaxies in our sample. 

We find that \fiddy\ has a large ICF(S) $=1.87\pm0.19$, leading to a higher $\log(\text{S/O}) = -1.49 \pm 0.07$. The other galaxies have S ionization corrections that are consistent with 1, making \fiddy\ a notable outlier. \fiddy\ also has the highest proportion of oxygen in the O$^{++}$ state, and the youngest stellar population with an age $<10$ Myr which is likely producing more high-energy photons. The \cite{izotov_chemical_2006} ICF for sulfur shows a significant upturn in this high-ionization range, as it is expected that the unobserved high-ionization S$^{+3}$ state will become more highly populated. However, other ionization corrections disagree as to the extent of this upturn. Notably, when using the ICF of \cite{amayo_ionization_2021}, we find an ICF(S)$= 1.28 \pm 0.29$, leading to a lower $\log(\text{S/O}) = -1.65 \pm 0.12$. Given the disagreement between these two ionization corrections and the significant systematic uncertainty implied by this disagreement, we interpret the total S abundance of \fiddy\ with caution.
%Given that both the ICFs of \cite{izotov_chemical_2006} and \cite{amayo_ionization_2021} are based on photoionization models selected to reproduce trends seen in local galaxies and \hii\ regions, it is possible that these are not accurate representations of young star-forming galaxies at higher redshift. Updated ICFs calibrated to match high-$z$ galaxies are needed, however the derivation of such ICFs is outside the scope of this work. We simply note that the total S abundance of \fiddy\ should be interpreted with caution given the variation observed between ionization corrections. 

%%%%%%%%%%%%%%%%%%%%%%%%%%%%%%%%%%%%%%%%%%%%%%%%%%%%%%%%%%%%%%
\section{Galaxy-Integrated Abundance Trends} \label{sec:discussion}

\begin{figure*}
    \centering
    \includegraphics[width=\linewidth]{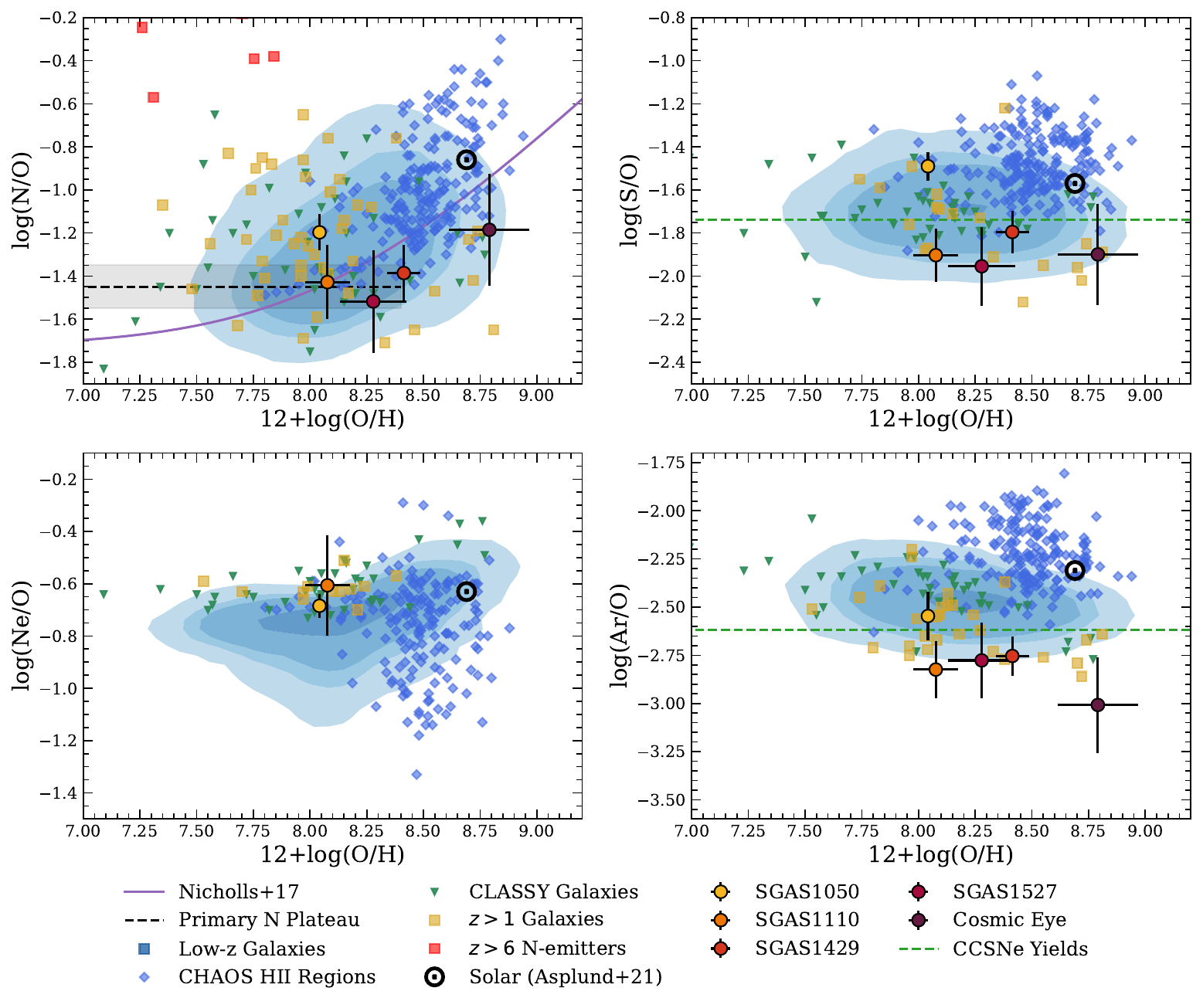}
    \caption{Abundance ratios of N/O (top left), S/O (top right), Ne/O (bottom left), and Ar/O (bottom right) are shown as a function of $12+\log(\text{O/H})$ for the LEGGOS galaxies (circles with errorbars). We also plot local \hii\ regions from the CHAOS survey (dark blue diamonds), $z\sim0$ galaxies from SDSS and DESI (blue contours) and CLASSY (green triangles), $z>1$ galaxies observed with JWST (yellow squares), and several $z>6$ UV N-emitters (red squares) as comparison samples (see Section \ref{sec:discussion} for details). Solar abundances from \cite{asplund_chemical_2021} are marked with a black $\odot$. We plot the \cite{nicholls_abundance_2017} N/O relation derived from Milky Way stars as a purple line in the top left panel. We additionally plot the IMF-averaged S/O and Ar/O ratios from CCSNe yields from \cite{kobayashi_galactic_2006} as a green dashed line in the top-right and bottom-right panels. The five lensed galaxies presented in this work are consistent with local galaxies and \hii~regions in both N/O and Ne/O. Similar to other high-$z$ galaxies, our sample mostly shows underabundances in S/O and Ar/O, suggesting enrichment from CCSNe. }
    \label{fig:abunds}
\end{figure*}

The abundance ratios of N, S, Ne, and Ar relative to O for our galaxy sample are plotted against the oxygen abundance in Figure \ref{fig:abunds}, alongside several literature comparison samples with direct $T_e$ abundance measurements. We include a large sample of $>50,000$ star-forming galaxies at low-$z$ from both SDSS \citep{izotov_chemical_2006} and DESI \citep{scholte_electron_2026}, as well as a sample of \hii\ regions from nearby spiral galaxies from the CHAOS program \citep{berg_chaos_2015,croxall_chaos_2015,croxall_chaos_2016,berg_chaos_2020,rogers_chaos_2021}. We additionally show a sample of $z\sim 0$ galaxies from the CLASSY program, which were selected as high-$z$ galaxy analogs \citep{berg_cos_2022,arellano-cordova_classy_2024,arellano-cordova_classy_2025}, a collection of $z>1$ galaxies with direct $T_e$ abundances from JWST observations \citep{rogers_cecilia_2026,arellano-cordova_jwst_2025,stanton_jwst_2025,cameron_jades_2026,cataldi_marta_2025,curti_marta_2025, welch_templates_2024,welch_sunburst_2025}, and several examples of $z>6$ galaxies with strong UV N emission \citep{senchyna_gn-z11_2024,castellano_jwst_2024,marques-chaves_extreme_2024,naidu_cosmic_2026,berg_fleeting_2026}. Here, we discuss the abundance trends observed for our lensed galaxy sample relative to other local and high-$z$ observations.

\subsection{Nitrogen}

The galaxy sample presented here generally matches nitrogen enrichment trends established in low-$z$ galaxies. \elf, \geese, and \batleth\ are consistent within uncertainties of the primary N plateau $\log(\text{N/O}) \simeq -1.45$ observed locally \citep{van_zee_oxygen_2006,scholte_electron_2026}. \elf, \geese, and \batleth\ each have slightly higher $12+\log(\text{O/H})$ than is typical of local galaxies on the primary N plateau; however, they all show SFHs with recent star formation within the last 100 Myr prior to observation. Secondary N enrichment is generally dominated by the asymptotic giant branch phase of low- to intermediate-mass stars (LIMS) \citep{henry_cosmic_2000,vincenzo_nitrogen_2016}.  It is therefore plausible that these galaxies have experienced O enrichment from CCSNe while the N enrichment has yet to catch up. \eye\ is slightly elevated in N/O, though this is in line with expectations given that this galaxy also has a super-solar oxygen abundance (see Table \ref{tab:results}). However, \eye\ is also on the lower end of the trend observed in local high-metallicity \hii\ regions, albeit with larger uncertainties. Given that this galaxy also went through a burst of star formation within the last 100 Myr, it is plausible that this galaxy is also observed between the O enrichment from CCSNe and the N enrichment from LIMS.

Other studies of high-$z$ galaxies have found mixed results for population-wide N/O enhancement, with some finding no evidence of enhancement \citep{schaerer_nitrogen_2026}, while others do find evidence for systematically enhanced N/O \citep{cataldi_tracing_2025,cameron_jades_2026}. Our sample matches the lack of general N/O enhancement seen in \cite{schaerer_nitrogen_2026}, although the small sample size of 5 galaxies means that we do not have the statistical weight to properly investigate average N enhancement in Cosmic Noon galaxies. 

%Observations of low-$z$ galaxies typically find no correlation between N/O and O/H at low oxygen abundance ($12+\log(\text{O/H}) \lesssim 8.2$), while at higher O/H the N/O increases rapidly \citep[e.g.,][]{van_zee_oxygen_2006,izotov_chemical_2006,scholte_electron_2026}. However some recent observations of high-z galaxies have reported excess N abundances at low O/H, with the most extreme examples being high-ionization N\textsc{iii}] and N\textsc{iv}] emitters at high-$z$ \citep[e.g.,][]{senchyna_gn-z11_2024,castellano_jwst_2024,marques-chaves_extreme_2024,martinez_under_2025,berg_fleeting_2026,morel_discovery_2025}. Outside of these extreme examples, results at Cosmic Noon have been somewhat mixed, with some authors finding general enhancement of N/O at low O/H \citep[e.g.,][]{cataldi_tracing_2025,cameron_jades_2026}, while other studies observe larger scatter \citep{rogers_cecilia_2026}. Meanwhile \cite{schaerer_nitrogen_2026} recently examined a sample of Lyman-continuum leaking galaxies at $z\sim 3$ and found that they exhibited similar N/O abundance patterns as LyC leakers at low redshift \citep{izotov_abundances_2023} and are not significantly enhanced relative to other low-$z$ star-forming galaxies. 

\fiddy\ is slightly above the primary N plateau with $\log(\text{N/O}) = -1.20 \pm 0.08$ despite its lower metallicity $12+\log(\text{O/H}) = 8.04\pm 0.01$, though it is still fully consistent with other local star-forming galaxies at similar metallicity \citep[e.g.,][]{arellano-cordova_classy_2025}. Because \fiddy\ resembles the N enhancement seen in some samples of high-$z$ galaxies \citep{cataldi_tracing_2025,cameron_jades_2026}, we next discuss the possible drivers of this mild N/O enhancement.

\fiddy\ is the youngest galaxy in our sample, with the bulk of its star formation occurring in a burst within the last 10 Myr. The moderately elevated N abundance is inconsistent with enrichment from LIMS, as these stars require a minimum of 100 Myr before their nucleosynthetic products enrich the surrounding medium \citep{kobayashi_origin_2020}. The SFH is also inconsistent with LIMS enrichment from a previous burst, as no earlier epoch shows any significant star formation activity (see Figure \ref{fig:sedfits1}). Notably, \fiddy\ shows a clear detection of nebular \heiiopt, a high-ionization emission line (I.P. 54.9 eV) which suggests the presence of a harder ionizing source than is present in typical stellar populations. The exact source of nebular \heii\ emission remains unclear \citep[e.g.,][]{olivier_characterizing_2022}, though proposed sources include WR stars \citep{shirazi_strongly_2012}, high-mass x-ray binaries \citep{schaerer_x-ray_2019}, and metal-poor massive stars, including VMS \citep{kehrig_extended_2015,kehrig_extended_2018,senchyna_high-mass_2020,kehrig_contribution_2021}. In low-$z$ galaxies, detections of \heiiopt\ driven by WR stars appear correlated with N enhancement, both in large samples of galaxies \citep{brinchmann_galaxies_2008} and within individual galaxies \citep[e.g.,][]{james_vlt_2009,pruijt_nitrogen_2026}. Other high-$z$ \heiiopt\ emitters have been suggested to have elevated N abundances driven by either WR stars or VMS \citep{rivera-thorsen_sunburst_2024,welch_sunburst_2025,curti_marta_2025,berg_fleeting_2026}, enrichment from VMS has been suggested to explain the N abundance of GNz11, which shows weak nebular \heiiuv\ emission \citep{bunker_jades_2023,vink_very_2023}. It is possible that \fiddy\ contains either WR stars or metal-poor (very) massive stars driving a temporarily elevated N/O abundance. Because N-enhancement from WR stars or VMS is expected to be a short-lived state, a better characterization of the stellar population age of \fiddy\ would be required to further investigate these massive stars as a potential source of N enrichment. % N enhancements driven by these massive stellar populations can be a local phenomenon, occurring at the level of an individual \hii\ region or single star cluster, as is the case for the Sunburst Arc \citep{pascale_nitrogen-enriched_2023,rivera-thorsen_sunburst_2024,welch_sunburst_2025}. 

\subsection{$\alpha$-elements: Neon, Sulfur, and Argon}

The $\alpha$-elements Ne, S, and Ar are expected to enrich simultaneously with O, given that they are generally produced by the same processes in CCSNe \citep[e.g.,][]{kobayashi_nucleosynthesis_2025}. This has been corroborated with local observations, which find nearly flat correlations between Ne/O, S/O, and Ar/O as a function of O/H \citep{berg_chaos_2020,rogers_chaos_2021,arellano-cordova_classy_2024,scholte_electron_2026}. 

In high-$z$ galaxies, Ne abundances have generally been observed to fluctuate around the solar Ne/O ratio, similar to local star-forming galaxies and predictions from galactic chemical enrichment models \citep{stanton_jwst_2025,welch_sunburst_2025}. This same result has also been observed in local high-$z$ analogs \citep{arellano-cordova_classy_2024}. However, some studies have derived Ne/O ratios that are significantly sub-solar, and have suggested that CCSNe from massive, metal-poor progenitors could drive this low Ne abundance \citep[e.g.,][]{isobe_jwst_2023}. We find that the two galaxies in our sample with measured Ne abundances have Ne/O consistent with solar within $1\sigma$ uncertainties, matching standard Ne production models \citep[e.g.,][]{kobayashi_origin_2020}. Our result is dependent on the choice of atomic data; when we use the collision strengths calculated by \cite{mclaughlin_large-scale_2011}, we find Ne/O consistent with solar, while using the earlier collision strengths of \cite{mclaughlin_electron_2000} yields systematically lower Ne/O, around $\sim 0.1$ dex lower in both \fiddy\ and \elf.

Argon and sulfur in high-z galaxies are the subject of an increasing body of literature, as they often appear underabundant relative to low-z galaxies of similar metallicity \citep{stanton_jwst_2025,bhattacharya_unveiling_2025,rogers_cecilia_2026}. In contrast, many studies of low-$z$ galaxies find Ar/O and S/O ratios that are relatively flat as a function of metallicity \citep{izotov_chemical_2006,arellano-cordova_classy_2024,scholte_electron_2026}, though these samples do contain several objects with underabundances of Ar/O and S/O. The difference is generally attributed to enrichment from CCSNe with no major SNIa contribution. SNIa can produce around 29\% of the S and 34\% of the Ar in galaxy chemical evolution models, but require longer timescales, typically $\sim 1$ Gyr, for this enrichment to occur \citep{kobayashi_origin_2020}. 

Our galaxy sample generally follows the trends seen in other high-z studies, with Ar/O and S/O underabundant relative to local star-forming galaxies and \hii\ regions. In the right-hand panels of Figure \ref{fig:abunds}, we plot the expected yields from CCSNe derived from \cite{kobayashi_galactic_2006} alongside our measured gas-phase abundances. The galaxies presented in this work are generally more consistent with the CCSNe yields than the solar abundance in both S/O and Ar/O, with the exception of \fiddy\ having S/O slightly above the solar value. As discussed in Section \ref{sec:physconditions}, the S/O abundance of \fiddy\ may be biased high by the ICF from \cite{izotov_chemical_2006}, and using the alternate ICF of \cite{amayo_ionization_2021} yields $\log(\text{S/O}) = -1.65 \pm 0.1$, consistent with the CCSNe yields. 

The SFHs of our sample provide additional support for the interpretation of enrichment by CCSNe. Each galaxy has undergone a recent epoch of star formation lasting less than 100 Myr. This is significantly shorter than the timescales required for SNIa, which are typically $\sim 0.1 - 1$ Gyr for main-sequence + white dwarf SNIa \citep{kobayashi_origin_2020}. We would not expect significant enrichment from SNIa from the most recent burst of star formation in each of these sources.  

Four of the five galaxies in our sample have S/O and Ar/O abundances that are below the predicted CCSNe yields of \cite{kobayashi_origin_2020}, though none of our galaxies has either S/O or Ar/O more than $2\sigma$ below the CCSNe yields. While it is possible that this underabundance relative to the predicted yields is simply scatter due to measurement uncertainty, it is important to note that the \cite{kobayashi_origin_2020} chemical enrichment model is specifically tailored to the Milky Way. The galaxies in our sample have different SFHs which could change the total yields of O, S, and Ar. Future studies could attempt to create bespoke enrichment models for any given star formation history, but such an endeavor is beyond the scope of this paper.

%%%%%%%%%%%%%%%%%%%%%%%%%%%%%%%%%%%%%%%%%%%%%%%%%%%%%%%%%%%%%%
\section{Conclusions} \label{sec:conclusions}

In this paper, we have derived direct $T_e$ abundances of N, O, Ne, S, and Ar for five strongly lensed galaxies in the LEGGOS sample \citep{khullar_leggos_2026}. 

We measure three $T_e$-sensitive auroral lines across the three ionization zones in two galaxies from our sample, plus two $T_e$ lines in a third galaxy. The temperatures measured are consistent with the scatter observed for both local and high-$z$ temperature scaling relations, as shown in Figure \ref{fig:temps}. 

We find a spread in oxygen abundances among these five galaxies, ranging from $8.04 \leq 12+\log(\text{O/H}) \leq 8.79$ (22\% to 126\% $Z_{\odot}$). We find that our sample generally matches the N/O-O/H trends seen in local star-forming galaxies, with minimal evidence for N enhancement. \fiddy\ shows some sign of elevated N/O, though it is still consistent with the scatter observed in low-$z$ galaxies. \fiddy\ also has multiple high-ionization emission lines in its spectra, and its SFH indicates it contains a very young stellar population with an age $<10$ Myr. We suggest that the mild elevation in N/O in \fiddy\ could be driven by short-term enrichment from massive stars (e.g., WR stars or VMS); however, additional investigation is required. 

We find that our galaxy sample trends towards low S/O and Ar/O, similar to other galaxies at $z>2$. These abundances are consistent with theoretical models of enrichment from CCSNe with a minimal contribution from SNeIa. Each galaxy has experienced significant star formation within $\sim100$ Myr prior to observation, matching expectations for CCSNe timescales while being too short for significant enrichment from SNeIa. 

We conclude that the LEGGOS galaxies are a sample of chemically-typical galaxies at Cosmic Noon that happen to be strongly magnified by gravitational lensing. Follow-up studies of these galaxies will investigate the spatial variations in physical conditions and chemical abundances within the galaxies to shed further light on the enrichment mechanisms shaping galaxies at the peak of cosmic star formation.

%%%%%%%%%%%%%%%%%%%%%%%%%%%%%%%%%%%%%%%%%%%%%%%%%%%%%%%%%%%%%%
\begin{acknowledgements}
      BW would like to acknowledge that the original name for the LEGGOS program was ``Clumps `n' Us". 

      We do not use the full name of JWST due to the person after whom this telescope is named and their role as NASA administrator during the ``Lavender Scare'', as per the $\#RenameJWST$ protest movement. GK would like to thank the Baum Grant and Fellowship at the University of Washington for support during this work, as well as the ALMA Ambassador Program (administered by NAASC and NRAO). GK would also like to thank the DiRAC Institute in the Department of Astronomy at the University of Washington. The DiRAC Institute is supported through generous gifts from the Charles and Lisa Simonyi Fund for Arts and Sciences, Janet and Lloyd Frink, and the Washington Research Foundation. GK is also grateful to the International Space Science Institute (ISSI), Bern, for their hospitality, financial support and collaboration during the time of writing this manuscript.
      
      This work is based primarily on observations made with the NASA/ESA/CSA \emph{JWST}. The data were obtained from the Mikulski Archive for Space Telescopes at the Space Telescope Science Institute, which is operated by the Association of Universities for Research in Astronomy, Inc., under NASA contract NAS 5-03127 for JWST. These observations are associated with JWST Cycle 2 GO programs \#4125 and 3843. The specific observations analyzed can be accessed via \href{https://doi.org/10.17909/70dh-0x38}{https://doi.org/10.17909/70dh-0x38} (GO-4124) and \href{https://doi.org/10.17909/h90f-n539}{https://doi.org/10.17909/h90f-n539} (GO-3843). Support for the GO programs was provided by NASA through a grant from the Space Telescope Science Institute, which is operated by the Associations of Universities for Research in Astronomy, Incorporated, under NASA contract NAS5-26555.
\end{acknowledgements}

\bibliographystyle{aa} % style aa.bst
\bibliography{zotero} % your references Yourfile.bib

%%%%%%%%%%%%%%%%%%%%%%%%%%%%%%%%%%%%%%%%%%%%%%%%%%%%%%%%%%%%%%%
% Appendices must be placed after   \end{thebibliography}
% They will be placed automatically on a new page.
%%%%%%%%%%%%%%%%%%%%%%%%%%%%%%%%%%%%%%%%%%%%%%%%%%%%%%%%%%%%%%%
\begin{appendix}

\onecolumn
\section{Table of emission line fluxes for LEGGOS spatially-integrated spectra}

\begin{table*}[ht!]

\caption {Emission Line Intensities}
\label{table:flux} 
\centering
\begin{tabular}{ccccccc}
\hline\hline
 \multicolumn{7}{c}{$I(\lambda)/I(H\beta)$} \\ 
\hline
Line & Wavelength (\AA) & \fiddy & \eye & \batleth & \geese & \elf \\
\hline \\[-1.5ex]
OIII]1666\tablefootmark{a} & $1666.15$ &$0.184 \pm 0.029$ & ... & ... & ... & ...  \\

NIII]1750 & $1750.00$ &$0.151 \pm 0.027$ & ... & ... & ... & ...  \\ 

SiIII] + CIII]\tablefootmark{b} & $1908.73$ &$0.598 \pm 0.021$ & ... & ... & ... & ...  \\ 

[OII]3727 & $3727.09$ &$0.556 \pm 0.020$ & ... & ... & ... & $0.895 \pm 0.031$  \\ 

[OII]3729 & $3729.88$ &$0.670 \pm 0.020$ & $4.222 \pm 0.865$ & $2.636 \pm 0.265$ & $2.389 \pm 0.218$ & $1.288 \pm 0.031$  \\ 

[NeIII]3870 & $3869.86$ &$0.534 \pm 0.004$ & ... & ... & ... & $0.494 \pm 0.028$  \\ 

HeI3889 & $3889.75$ &$0.178 \pm 0.004$ & ... & ... & ... & $0.304 \pm 0.041$  \\ 

[NeIII]3969 & $3968.59$ &$0.144 \pm 0.018$ & ... & ... & ... & $0.132 \pm 0.055$  \\ 

Hepsilon & $3971.20$ &$0.189 \pm 0.018$ & ... & ... & ... & $0.114 \pm 0.043$  \\ 

Hdelta & $4102.89$ &$0.266 \pm 0.003$ & $0.267 \pm 0.042$ & ... & ... & $0.231 \pm 0.027$  \\ 

Hgamma & $4341.68$ &$0.485 \pm 0.003$ & $0.492 \pm 0.020$ & $0.297 \pm 0.011$ & ... & $0.473 \pm 0.029$  \\ 

[OIII]4363 & $4364.44$ &$0.122 \pm 0.002$ & ... & $0.037 \pm 0.007$ & ... & $0.113 \pm 0.041$  \\ 

HeI4473 & $4472.70$ &$0.038 \pm 0.002$ & ... & ... & ... & ...  \\ 

HeII4687 & $4687.02$ &$0.017 \pm 0.002$ & ... & ... & ... & ...  \\ 

[ArIV]4713 & $4712.69$ &$0.008 \pm 0.001$ & ... & ... & ... & ...  \\ 

HeI4714 & $4714.47$ &$0.005 \pm 0.000$ & ... & ... & ... & ...  \\ 

[ArIV]4741 & $4741.45$ &$0.007 \pm 0.002$ & ... & ... & ... & ...  \\ 

Hbeta & $4862.68$ &$1.000 \pm 0.002$ & $1.000 \pm 0.009$ & $1.000 \pm 0.004$ & $1.000 \pm 0.013$ & $1.000 \pm 0.013$  \\ 

[OIII]4960 & $4960.30$ &$2.417 \pm 0.002$ & $0.891 \pm 0.005$ & $1.430 \pm 0.004$ & $1.802 \pm 0.009$ & $1.792 \pm 0.013$  \\ 

[OIII]5008 & $5008.24$ &$7.441 \pm 0.004$ & $2.729 \pm 0.006$ & $4.177 \pm 0.005$ & $5.152 \pm 0.013$ & $5.587 \pm 0.016$  \\ 

[NII]5756 & $5756.24$ &$0.008 \pm 0.002$ & ... & ... & ... & ...  \\ 

HeI5877 & $5877.25$ &$0.122 \pm 0.001$ & $0.138 \pm 0.002$ & $0.130 \pm 0.002$ & $0.114 \pm 0.005$ & $0.131 \pm 0.006$  \\ 

[OI]6302 & $6302.05$ &$0.041 \pm 0.001$ & $0.086 \pm 0.002$ & $0.071 \pm 0.002$ & $0.058 \pm 0.005$ & $0.051 \pm 0.005$  \\ 

[SIII]6314 & $6313.80$ &$0.011 \pm 0.001$ & ... & $0.012 \pm 0.002$ & $0.010 \pm 0.005$ & $0.021 \pm 0.004$  \\ 

[OI]6366 & $6365.54$ &... & $0.018 \pm 0.002$ & $0.020 \pm 0.002$ & $0.023 \pm 0.005$ & ...  \\ 

[NII]6550 & $6549.85$ &$0.043 \pm 0.001$ & $0.263 \pm 0.002$ & $0.096 \pm 0.002$ & $0.017 \pm 0.004$ & $0.020 \pm 0.004$  \\ 

Halpha & $6564.61$ &$2.818 \pm 0.002$ & $2.819 \pm 0.003$ & $2.818 \pm 0.003$ & $2.818 \pm 0.007$ & $2.818 \pm 0.007$  \\ 

[NII]6585 & $6585.28$ &$0.087 \pm 0.001$ & $0.779 \pm 0.002$ & $0.235 \pm 0.002$ & $0.111 \pm 0.004$ & $0.108 \pm 0.004$  \\ 

HeI6680 & $6680.00$ &$0.024 \pm 0.001$ & $0.027 \pm 0.002$ & $0.027 \pm 0.002$ & $0.014 \pm 0.009$ & $0.025 \pm 0.004$  \\ 

[SII]6718 & $6718.29$ &$0.080 \pm 0.001$ & $0.240 \pm 0.002$ & $0.186 \pm 0.002$ & $0.146 \pm 0.004$ & $0.152 \pm 0.004$  \\ 

[SII]6733 & $6732.67$ &$0.066 \pm 0.001$ & $0.225 \pm 0.002$ & $0.155 \pm 0.002$ & $0.113 \pm 0.004$ & $0.086 \pm 0.004$  \\ 

HeI7067 & $7067.16$ &$0.050 \pm 0.001$ & $0.019 \pm 0.002$ & $0.023 \pm 0.002$ & $0.018 \pm 0.004$ & $0.023 \pm 0.004$  \\ 

HeI7283 & $7283.36$ &$0.010 \pm 0.002$ & ... & ... & ... & ...  \\ 

[ArIII]7138 & $7137.80$ &$0.054 \pm 0.001$ & $0.041 \pm 0.002$ & $0.061 \pm 0.002$ & $0.046 \pm 0.003$ & $0.038 \pm 0.004$  \\ 

[OII]7322 & $7322.01$ &... & $0.041 \pm 0.002$ & $0.025 \pm 0.002$ & $0.029 \pm 0.004$ & $0.020 \pm 0.005$  \\ 

[OII]7332 & $7331.68$ &$0.047 \pm 0.002$ & $0.025 \pm 0.002$ & $0.020 \pm 0.002$ & $0.022 \pm 0.005$ & $0.030 \pm 0.003$  \\ 

[ArIII]7753 & $7753.20$ &... & ... & $0.016 \pm 0.002$ & ... & ...  \\ 

OI8449 & $8449.08$ &$0.017 \pm 0.001$ & ... & ... & ... & ...  \\ 

Pa10 & $9017.39$ &$0.015 \pm 0.001$ & ... & $0.004 \pm 0.002$ & ... & ...  \\ 

[SIII]9071 & $9071.10$ &$0.127 \pm 0.001$ & $0.206 \pm 0.002$ & $0.184 \pm 0.002$ & $0.113 \pm 0.004$ & $0.127 \pm 0.004$  \\ 

Pa9 & $9231.55$ &$0.027 \pm 0.001$ & ... & $0.020 \pm 0.001$ & ... & ...  \\ 

[SIII]9533 & $9533.20$ &$0.341 \pm 0.001$ & $0.447 \pm 0.002$ & $0.457 \pm 0.001$ & $0.278 \pm 0.003$ & $0.330 \pm 0.003$  \\ 

Paepsilon & $9548.59$ &... & ... & ... & ... & ...  \\ 

Padelta & $10052.13$ &$0.066 \pm 0.001$ & $0.025 \pm 0.001$ & $0.037 \pm 0.001$ & $0.036 \pm 0.003$ & $0.037 \pm 0.004$  \\ 

Pagamma & $10941.09$ &$0.109 \pm 0.001$ & $0.062 \pm 0.001$ & $0.093 \pm 0.001$ & $0.095 \pm 0.004$ & $0.086 \pm 0.004$  \\ 

HeI10833 & $10833.22$ &$0.457 \pm 0.001$ & $0.303 \pm 0.001$ & $0.314 \pm 0.001$ & $0.302 \pm 0.003$ & $0.262 \pm 0.003$  \\ 

[FeII]12570 & $12570.24$ &... & $0.030 \pm 0.001$ & $0.022 \pm 0.001$ & $0.018 \pm 0.003$ & ...  \\ 

Pabeta & $12821.59$ &... & $0.190 \pm 0.001$ & $0.191 \pm 0.001$ & $0.147 \pm 0.003$ & $0.161 \pm 0.003$  \\  
\hline \\[-1.5ex]
$F(H\beta)$ & & $5.33 \pm 0.010$ & $2.83 \pm 0.025$ & $4.15 \pm 0.016$ & $1.46 \pm 0.019$ & $1.19 \pm 0.015$ \\
$E(B-V)$ & & $0.1190 \pm 0.0005$ & $0.268 \pm 0.002$ & $0.164 \pm 0.001$ & $0.065 \pm 0.004$ & $0.097 \pm 0.004$ \\
\hline
\end{tabular}
\tablefoot{Measured emission line intensities relative to \hbeta\ for the galaxies presented in this work. The observed image-plane flux of \hbeta\ is given for each target in units of $10^{-16}\text{erg s}^{-1}\text{ cm}^{-2}$. The He\textsc{i}$\lambda4714$ intensity for \fiddy\ is calculated from the nearby He\textsc{i}$\lambda4473$ line. Note that the \oiibrightdbl\ lines are blended in the medium-resolution spectrum of \fiddy. In cases where neighboring doublets are heavily blended, the total flux is given as the redder line flux. \\
\tablefoottext{a}{Blend of the O\textsc{iii}]$\lambda\lambda$1660,1666 doublet and the He\textsc{ii}$\lambda1640$ line in the prism spectrum.} \\
\tablefoottext{b}{Blend of the \siliiidbl\ and \ciiidbl\ lines in the prism spectrum.}
}
\label{tab:fluxes}
\end{table*}

\end{appendix}
\end{document}